\documentstyle[12pt,aaspp4]{article}

\begin{document}

\title{An Efficient Technique to Determine the Power Spectrum from Cosmic
Microwave Background Sky Maps}
\author{Siang Peng Oh, David N. Spergel}
\affil{Princeton University Observatory, NJ 08544} 

\and

\author{Gary Hinshaw}
\affil{Goddard Space Flight Center, MD 20771}

\begin{abstract}
There is enormous potential to advance cosmology from statistical 
characterizations of cosmic microwave background sky maps.  The angular power 
spectrum of the microwave anisotropy is a particularly important statistic.
Existing algorithms for computing the angular power spectrum of a pixelized 
map typically require $O(N^3)$ operations and $O(N^2)$ storage, where  $N$ is
the number of independent pixels in the map. The {\it MAP} and {\it Planck} 
satellites will produce megapixel maps of the cosmic microwave background 
temperature at multiple frequencies; thus, existing algorithms are not 
computationally feasible. In this article, we introduce an algorithm that 
requires $O(N^2)$ operations and $O(N^{3/2})$ storage that can find the 
minimum variance power spectrum from sky map data roughly one million times 
faster than was previously possible.  This makes feasible an analysis that 
was hitherto intractable.
\end{abstract}

\section{Introduction}

The {\it MAP} and {\it Planck} satellites have the potential to yield enormous 
amounts of information about the physical conditions in the early universe 
(Bennett et al. 1995; http://map.gsfc.nasa.gov; Bersanelli et al. 1996; 
http://astro.estec.esa.nl/Planck). For example, with full sky coverage and 
an angular resolution of $0.21^{\circ }$ at its highest frequency, {\it MAP} 
should return accurate measurements of the cosmic microwave background (CMB) 
temperature anisotropy at roughly one million independent points on the sky. 
If the temperature fluctuations are consistent with inflationary models, these
measurements will provide accurate determinations of most of the significant
cosmological parameters (Spergel 1994; Knox 1995; Hinshaw, Bennett \& Kogut
1995; Jungman et al. 1996; Zaldarriaga, Spergel \& Seljak 1997; Bond,
Efstathiou \& Tegmark 1997). If they are not consistent with inflationary
models, it will be even more exciting as we will need to rethink our ideas
about the physics of the early universe.

There are a number of non-trivial numerical steps involved in comparing 
$10^{11}$ temperature differences with the predictions of a particular
cosmological model. Several of the steps in the calculation are now clear.
Wright, Hinshaw \& Bennett (1996) found a rapid and exact algorithm for
producing megapixel cosmic microwave background maps from differential data.
Seljak \& Zaldarriaga (1996) present a fast algorithm for computing the power
spectrum for a given cosmological model. However, to date, we have lacked an
efficient technique for computing the power spectrum from an observed sky map.
The techniques that were used to extract the power spectrum from the {\it COBE}
data (G\'orski 1994; G\'orski et al. 1994; Bond 1995; Tegmark \& Bunn 1995; 
G\'orski et al. 1996; Hinshaw et al. 1996; Tegmark 1996; Bunn \& White
1997) cannot easily be 
extended to the high resolution data since these techniques require $O(N^3)$ 
operations and O($N^2$) storage. For a single frequency of {\it MAP} data, for
example, this would take $>10^{19}$ operations, with $>10^{21}$ operations 
required for a combined analysis of polarization and temperature
maps. All proposed techniques to date are thus wholly unequal to the
task of analysing upcoming data sets (Borrill 1997). 

In this paper, we present a fast, accurate method for extracting the power
spectrum from realistic simulations of high resolution data at a single 
frequency. In \S 2-5 we develop our numerical method for determining the power 
spectrum from data containing only cosmological signal and detector noise. 
In \S 6, we show how the same approach can be used to directly estimate 
cosmological parameters from the maps.  In \S 7, we apply the method to 
realistic simulations of {\it MAP} data that include spatially varying noise 
and a galactic sky cut.  We show that our numerical method recovers minimum 
variance, unbiased estimates of the power spectrum and of cosmological 
parameters.  We summarize our results in \S 8.  A subsequent paper will 
extend the techniques developed here to the analysis of multi-frequency data 
and polarization data.

\section{Maximum Likelihood Determination of the Power Spectrum}

The basic problem is to extract the angular power spectrum of the CMB 
temperature fluctuations from noisy data.  We can expand the CMB temperature 
in spherical harmonics,
\begin{equation}
\delta t(\Omega) = \sum_{l,m} a_{lm}^{sky} \, w_l \, Y_{lm}(\Omega)
\label{Expansion}
\end{equation}
where $w_l$ is Legendre expansion of the experimental window function. If 
inflationary models are correct, the $a_{lm}^{sky}$ coefficients are
uncorrelated Gaussian random variables with zero mean and a variance that is
independent of orientation
\begin{equation}
\left\langle a_{lm}^{sky}a_{l^{\prime}m^{\prime}}^{sky*} \right\rangle 
= c_l^{sky} \, \delta_{ll^{\prime}} \, \delta_{mm^{\prime}}
\end{equation}
where $c_l^{sky}$ is the angular power spectrum we wish to estimate from the 
maps.

Like {\it COBE}, {\it MAP} and {\it Planck} will  produce full sky maps at a 
number of frequencies.  In this paper, we will  focus on the analysis of the 
highest resolution map from {\it MAP} (0.21$^{\circ}$ at 94 GHz), which is 
expected to contain useful signal up to at least multipole order $l \sim 1000$. 
The data in a given map may be expressed as a superposition of several terms
\begin{equation}
{\bf m} = {\bf \delta t} + {\bf n} + {\bf g} + {\bf f}
\label{measure}
\end{equation}
where ${\bf \delta t}$ is the CMB signal convolved with the instrument's beam
response, ${\bf n}$ is the detector noise, ${\bf g}$ is the systematic error, 
and ${\bf f}$ is the foreground emission.  Note that the measurement  vector
${\bf m}$ can be expressed either in a pixel basis, with components $m_i$, the
observed temperature in pixel $i$, or in a spherical harmonic basis, with
components $m_{lm} = a_{lm}$, the observed moment of the spherical harmonic 
$Y_{lm}$.  The choice of basis to use for power spectrum estimation requires 
weighing many trade-offs which we discuss in this and subsequent sections.  
Key among these is the form of the covariance matrix that describes the data, 
${\bf m}$.

While {\it MAP} will measure temperature fluctuations across the entire sky, 
we will want to exclude pixels in the Galactic plane from our analysis, as 
well as other randomly located pixels that contain bright, non-cosmological 
sources.  In this paper, we will assume that we can remove foreground emission 
by simply excluding the galactic plane and bright sources.  The resulting 
incomplete coverage of the celestial sphere precludes naively evaluating a 
spherical harmonic expansion of the data by direct integration.  The analysis
is further complicated  by {\it MAP}'s inhomogeneous sampling of the sky: the
noise per unit area will  vary by more than a factor two between the ecliptic
poles and the equator.   {\it MAP}'s differential design has been optimized to
minimize systematic  errors and to produce maps with spatially uncorrelated
noise.  We will assume that we can ignore systematic errors and that the noise
is uncorrelated from pixel to pixel.  We have not yet tested the performance of
our algorithm with maps that that contain significant pixel to pixel noise
correlations or ``stripes''.

With the above assumptions, the covariance matrix of the data takes the form 
${\bf C} \equiv \langle {\bf m}{\bf m}^T \rangle = {\bf S} + {\bf N}$  where 
${\bf S} = \langle {\bf \delta t} {\bf \delta t}^T \rangle$ is the covariance 
of the CMB fluctuations, and ${\bf N} = \langle {\bf n}{\bf n}^T \rangle$ is 
the covariance matrix of the noise.  In the pixel basis, the noise matrix is 
diagonal (${\bf N}_{ij} = \sigma _i^2 \delta _{ij}$, where $\sigma_i$ is the 
$rms$ noise in pixel $i$) while in the spherical harmonic basis the 
signal matrix is diagonal
\begin{equation}
{\bf S}_{(lm)(lm)^{\prime}} \equiv 
\left\langle a_{lm}^{sky}a_{l^{\prime}m^{\prime}}^{sky*} \right\rangle 
w_l w_{l^{\prime}} = c_l^{sky} \, w_l^2 \, \delta_{(lm)(lm)^{\prime}}.
\label{signal}
\end{equation}
For the sake of notational simplicity, we shall henceforth set
\begin{equation}
c_l = c_l^{sky} \, w_l^2
\end{equation}
and solve for $c_l$. Note that the signal matrix in the pixel basis can be easily expressed in
terms of spherical harmonics
\begin{equation}
{\bf S}_{ij} \equiv \left\langle \delta t_i \delta t_j \right\rangle 
= \sum_{(lm),(lm)^{\prime}} Y_{lm}(\Omega_i) {\bf S}_{(lm)(lm)^{\prime}} 
Y_{l^{\prime}m^{\prime}}^{*}(\Omega_j) \equiv {\bf YSY}^{*}
\label{pixelsignal}
\end{equation}
where ${\bf Y}^{*}$ is the complex conjugate of ${\bf Y}$.  The form of the 
noise matrix in the spherical harmonic basis is deferred to the next section.

We wish to estimate the angular power spectrum from the data by explicitly 
maximizing the likelihood function, which, for Gaussian fluctuations, has the
form
\begin{equation}
{\cal L}(c_l|{\bf m}) = \frac{\exp \left( -\frac 12{\bf m}^T{\bf C}^{-1}
{\bf m}\right) }{(2 \pi)^{N/2}({\det {\bf C})^{1/2}}}
\label{Likelihood}
\end{equation}
where ${\bf C}$ is the full covariance matrix.  A number of authors 
(G\'orski 1994; G\'orski et al. 1994; Bond 1995; Tegmark \& Bunn 1995; 
G\'orski et al. 1996; Hinshaw et al. 1996; Tegmark 1996; Bunn \& White
1997) have used this 
approach to extract the power spectrum from the {\it COBE} data.

We find the most likely power spectrum by expanding the likelihood function
as a Taylor series 
\begin{equation}
f\equiv -2\ln {\cal L}=\bar f+\sum_l\left. \frac{\partial f}{\partial c_l}%
\right| _{\bar {c_l}}\left( c_l-\bar {c_l}\right) +\sum_{l,l^{\prime }}\frac
12\left. \frac{\partial ^2f}{\partial c_l\partial c_{l^{\prime }}}\right|
_{\bar {c_l}}\left( c_l-\bar {c_l}\right) \left( c_{l^{\prime }}-\bar
{c_{l^{\prime }}}\right)  \label{Taylor}
\end{equation}
where the over-bar indicates the values at which $f$ is minimized.
Differentiating equation (\ref{Likelihood}) yields 
\begin{equation}
\frac{\partial f}{\partial c_l}=-\left( \frac
\partial {\partial c_l}\right) 2\ln {\cal L}=-{\bf m}^T{\bf C}^{-1}{\bf P}^l%
{\bf C}^{-1}{\bf m}+{\rm tr}({\bf C}^{-1}{\bf P}^l)  \label{Gradlike}
\end{equation}
where 
\begin{equation}
{\bf P}^l \equiv \frac{\partial {\bf C}}{\partial c_l}.
\end{equation}
In the spherical harmonic basis ${\bf P}^l$ is a diagonal matrix with entries 
0 or 1, while in the pixel basis, it is given by
\begin{equation}
{\bf P}^l_{ij} = \frac{(2l+1)}{4\pi} \, P_l(\cos \gamma_{ij})
\end{equation}
where $P_l$ is the Legendre polynomial of order $l$, and $\gamma_{ij}$ is the 
angle between pixels $i$ and $j$.  The second order term in the likelihood 
function is 
\begin{equation}
\frac 12 \frac{\partial ^2f}{\partial
c_l\partial c_{l^{\prime }}}={\bf m}^T{\bf C}^{-1}{\bf P}^l{\bf C}^{-1}{\bf P%
}^{l^{\prime }}{\bf C}^{-1}{\bf m}-\frac 12{\rm tr}({\bf C}^{-1}{\bf P}^l%
{\bf C}^{-1}{\bf P}^{l^{\prime }}).
\end{equation}
The expectation value of this quantity is the Fisher matrix 
\begin{equation}
{\bf F}_{ll^{\prime }}=\left\langle
-\left( \frac{\partial ^2}{\partial c_l\partial c_{l^{\prime }}}\right) \ln 
{\cal L}\right\rangle =\frac 12{\rm tr}({\bf C}^{-1}{\bf P}^l{\bf C}^{-1}%
{\bf P}^{l^{\prime }})  \label{Fisher}
\end{equation}
where we have used $\langle {\bf m}{\bf m}^T\rangle ={\bf C}$.

We can locate the most-likely power spectrum by finding where the derivative
of the likelihood function is zero. Let $c^{(0)}_l$ be a trial power
spectrum, then the derivative of the likelihood in the neighborhood of $%
c^{(0)}_l$ may be written 
\begin{equation}
\left.\frac{\partial f}{\partial c_l}\right|_{c_l} \simeq \left.\frac{%
\partial f}{\partial c_l}\right|_{c^{(0)}_l} + 2\,\sum_{l^{\prime}}{\bf F}%
_{ll^{\prime}} \left(c_{l^{\prime}} - c^{(0)}_{l^{\prime}}\right) = 0
\end{equation}
where we have approximated the second derivative by its expectation value.
This suggests solving for the most-likely power spectrum using a variable
metric method, 
\begin{equation}
c_l^{(n+1)} = c_l^{(n)} - \frac{1}{2}\sum_{l^{\prime}}{\bf F}%
^{-1}_{ll^{\prime}} \left.\frac{\partial f}{\partial c_{l^{\prime}}}%
\right|_{c^{(n)}_{l^{\prime}}}.
\label{Newton}
\end{equation}
Note that this is exactly the Newton-Raphson method for non-linear systems
of equations (Press et al. 1992), which is well known to converge
quadratically near the neighborhood of a root. Its only potential problem is
its poor global convergence properties, which can occur when the
approximation in equation (\ref{Taylor}) is not valid. In practice, we have
found the likelihood function is sufficiently well-behaved that this is
never a problem, even with extremely poor starting guesses. The
structureless nature of the likelihood function has also been noted by other
authors (Bond, Jaffe \& Knox 1998). This is easy to understand intuitively.  
As we demonstrate later the $c_{l}$ parameters are only weakly correlated 
with each other as the Fisher matrix is diagonally dominant. Thus, to first 
order, we are performing a series of one-parameter likelihood maximizations, 
with small corrections for couplings. In each dimension, the likelihood is 
very well approximated by a parabola (since the probability distribution for 
each $c_{l}$ is close to Gaussian). In particular, no local extrema exist, so 
there is no danger in employing Newton-Raphson.  This also explains the 
extremely rapid convergence of the method -- typically 3-4 iterations.

Our equations are essentially equivalent to those of Bond, 
Jaffe \& Knox (1998) (in addition, Tegmark (1997) obtained equivalent
results by considering a quadratic optimisation problem). However, we have recast them into a form that is 
computationally more tractable.  The key advance is in reducing all 
ostensibly $O(N^3)$ operations to $O(N^2)$ operations.  We highlight the 
crucial elements of this improvement below, and discuss them more fully in 
subsequent sections. 

$\bullet$ ${\bf C}^{-1}{\bf m}$ -- Rather than evaluating this expression 
directly using Choleski techniques, we solve it iteratively using conjugate
gradient techniques (Press et al. 1992, Barrett et al. 1994), which are
applicable to symmetric, positive definite linear systems.  A good
approximate inverse or preconditioner, $\widetilde{{\bf C}}^{-1}$, is
essential to reducing the number of iterations required.  We find such a
preconditioner by exploiting the approximate azimuthal symmetry of the
noise pattern on the sky.  This approach requires $O((N/20)^{3/2})$ memory 
for storage and takes $O((N/20)^{2})$ operations to compute. In addition,
each iteration of the conjugate gradient method involves performing a
matrix multiplication of the form ${\bf C z}$ where ${\bf z}$ is a vector. 
We speed this up by writing ${\bf C}$ as a convolution of diagonal matrices 
and spherical harmonic transforms.  By employing fast spherical harmonic 
transforms (Muciaccia, Natoli \& Vittorio 1998; Driscoll \& Healey 1994), we 
are able to reduce the cost of the matrix multiplication from $O(N^{2})$ to 
$O(N^{3/2})$. 

$\bullet$ tr$({\bf C}^{-1}{\bf P}^{l})$ -- We first compute this term
approximately by assuming azimuthal symmetry of the noise. We then
compute it exactly using Monte Carlo simulations of maps and
exploiting the fact that
\begin{equation}
\left\langle {\bf m}^T{\bf C}^{-1}{\bf P}^l{\bf C}^{-1}{\bf m} \right\rangle
= {\rm tr}({\bf C}^{-1}{\bf P}^l)
\end{equation}
where we have used $\langle {\bf m}{\bf m}^T \rangle = {\bf C}$. The errors
obtained by computing the trace term with Monte Carlo simulations rather than 
exactly are $\sqrt{1+1/N_{mc}}$ times larger than the minimum variance
errors, where $N_{mc}$ is the number of simulations used to compute the trace.
Thus, if we generate 100 simulations, our errors will be only 0.5\% larger 
than the minimum variance errors. The total cost of this step is 
$O(N_{mc}N_{iter}N^{3/2})$, where $N_{iter}$ is the number of iterations used 
to evaluate equation (\ref{Newton}).

$\bullet$ ${\bf F}$ -- Note that we only require an approximate second 
derivative to converge to the maximum of the likelihood function -- the 
solution is fully independent of ${\bf F}$. We therefore compute ${\bf F}$ 
approximately using the preconditioner $\widetilde{{\bf C}}^{-1}$ previously 
computed, which requires $O((N/20)^{2})$ operations. Once we obtain the maximum
likelihood $c_{l}$, we use Monte Carlo simulations to obtain their probability
distribution and hence their errors.

\section{Iterative Evaluation of the Term ${\bf C}^{-1}{\bf m}$}

A key step in our analysis is to iteratively (and rapidly) solve the linear
equation
\begin{equation}
{\bf Cz}={\bf m}
\label{weiner}
\end{equation}
using conjugate gradient techniques.  Note that this solution is only used in
the evaluation of terms of the form ${\bf m}^T{\bf C}^{-1}{\bf P}^l{\bf C}^{-1}
{\bf m}$.  Thus, we have complete freedom to obtain this solution in pixel 
space (where the data vector is simply the temperature map) or in spherical 
harmonic space (where the data vector is the least squares fit of spherical
harmonic coefficients to the temperature map). In pixel space, we can use the 
addition theorem for spherical harmonics to write
\begin{equation}
{\bf m}^T{\bf C}^{-1}{\bf P}^l{\bf C}^{-1}{\bf m} 
= \sum_{m=-l}^{l}{\bf m}^T{\bf C}^{-1}{\bf Y}^{*}_{lm}
  {\bf Y}_{lm}{\bf C}^{-1}{\bf m}
\end{equation}
which involves taking spherical harmonic transforms of the filtered, cut map,
${\bf C}^{-1}{\bf m}$.  In spherical harmonic space, ${\bf P}^{l}$ is simply a
diagonal matrix with ones or zeros along the diagonal, so the evaluation of
${\bf m}^T{\bf C}^{-1}{\bf P}^l{\bf C}^{-1}{\bf m}$ is even simpler.

Our choice of the appropriate space to work in is motivated by a second
consideration: the conjugate gradient technique requires that we compute an
appropriate preconditioner matrix.  Given a linear system ${\bf A x}= {\bf
b}$, where ${\bf A}$ is symmetric and positive definite, the preconditioner is a symmetric positive definite matrix $\widetilde{{\bf A}}$, such that
$\widetilde{{\bf A}}^{-1}{\bf A=I+R,}$ where the eigenvalues of ${\bf R}$ are 
all less than 1. The preconditioned conjugate gradient technique then 
solves the system 
\begin{equation}
\widetilde{{\bf A}}^{-1}{\bf A x}=\widetilde{{\bf A}}^{-1}{\bf b}
\end{equation}
by generating a series of search directions and improved iterates. 
Specifically, it generates a sequence of coupled recurrence relations for the 
residual vector ${\bf r}^{(i)} \equiv {\bf b - A x}^{(i)}$ and the search
direction ${\bf p}^{(i)}$
\begin{eqnarray}
{\bf r}^{(i)} & = & {\bf r}^{(i-1)}-\alpha_{i}{\bf A p}^{(i)} \\
{\bf p}^{(i)} & = & \widetilde{\bf A}^{-1}{\bf r}^{(i-1)} + \beta_{i-1}
{\bf p}^{(i-1)}
\end{eqnarray}
The scalar $\alpha_{i}$ is chosen to minimize the quadratic function
$f({\bf x})=({\bf x}^{(i)}-\hat{\bf x})^T{\bf A}({\bf x}^{(i)}-\hat{\bf x})$, 
where $\hat{\bf x}$ is the exact solution to ${\bf A x}= {\bf b}$
\begin{equation}
\alpha_{i} = \frac {{\bf r}^{(i-1)T}\widetilde{\bf A}^{-1}{\bf
r}^{(i-1)}} {{\bf p}^{(i)} {\bf A p}^{(i)}}.
\end{equation}
The scalar $\beta_{i-1}$ is chosen to ensure that the residuals are 
$\widetilde{\bf A}^{-1}$ orthogonal (i.e., ${\bf r}^{(i)}\widetilde{\bf A}^{-1}
{\bf r}^{(j)} = 0$ for $i \ne j$)
\begin{equation}
\beta_{i-1} = \frac {{\bf r}^{(i-1)T}\widetilde{\bf A}^{-1}{\bf r}^{(i-1)}}
                    {{\bf r}^{(i-2)T}\widetilde{\bf A}^{-1}{\bf r}^{(i-2)}}.
\end{equation}
In this manner the quadratic function $f({\bf x})$ of the improved iterate
\begin{equation}
{\bf x}^{(i)}={\bf x}^{(i-1)} + \alpha_{i} {\bf p}^{(i)}
\end{equation}
is minimized over the whole vector space of conjugate search directions already
taken, $\{{\bf p}_{1},{\bf p}_{2},...\}$. The routine is initialized by setting 
${\bf r}^{(0)}={\bf b}-{\bf A x}^{(0)}$ for some initial guess ${\bf x}^{(0)}$, 
and setting ${\bf p}^{(1)}=\widetilde{\bf A}^{-1}{\bf x}^{(0)}$. In general, 
the number of search directions required to span the vector space of possible 
solutions is $N$; the preconditioner reduces this by transforming the contours 
of $f({\bf x})$ to be as spherical as possible. To the extent that this is
achieved the number of independent directions to minimize over becomes very 
small and the conjugate gradient routine converges quadratically. More 
precisely, the number of iterations required is proportional to $\sqrt 
\kappa_{2}$, where $\kappa_{2}$ is the condition number of the matrix 
$\widetilde{\bf A}^{-1}{\bf A}$ (i.e., the ratio of its largest to smallest 
eigenvalue). Further details may be found in Barrett et al. (1994) and Press 
et al. (1992). 

There are two conflicting requirements for a preconditioner: it must be a
sufficiently good approximation to the true inverse that $\kappa_{2}$ is not 
too large, yet it must be sufficiently sparse that it is significantly easier 
to compute and store than the original matrix. From this comes our choice of 
the correct space to work in: at low $l$ where the signal dominates, we want 
to work in spherical harmonic space (where ${\bf S}$ is diagonal), while at 
high $l$, where the noise dominates, we want to work in pixel space (where 
${\bf N}$ is diagonal).

Thus, we begin by considering how to transform the system into spherical 
harmonic space.  We can obtain the best-fit multipole moments,
$m_{lm}=w_l\,m_{lm}^{sky}$, by minimizing
\begin{equation}
\chi^2 = \sum_i \frac{1}{\sigma_i^2}
\left[m_i - \sum_{l,m} m_{lm}\,Y_{lm}(\Omega_i)\right]^2
\label{Chisq}
\end{equation}
where the sum is over the uncut pixels, $\sigma_i$ is the $rms$ noise in the 
$i$th pixel, $m_i$ is the observed temperature of the $i$th pixel, $\Omega_i$ 
is the direction to the center of the $i$th pixel, and $w_l$ is the
Legendre transform of the experimental window function.  By differentiating
equation (\ref{Chisq}), we derive the normal equations (Press et al. 1992)
\begin{equation}
{\bf N}^{-1}{\bf m} = {\bf y}
\label{Normal}
\end{equation}
where
\begin{equation}
{\bf N}^{-1}_{(lm)(lm)^{\prime}} = \sum_i 
\frac{Y_{lm}(\Omega_i)\,Y_{l^{\prime}m^{\prime}}
(\Omega_i)}{\sigma_i^2}
\label{Noise}
\end{equation}
is the inverse of the noise matrix for the multipole moments, and
\begin{equation}
{\bf y}_{(lm)} = \sum_i \frac{m_i\,Y_{lm}(\Omega_i)}{\sigma_i^2}
\label{tlm}
\end{equation}
is the variance weighted spherical harmonic transform of the temperature map.
The solution of the set of normal equations is a minimum variance estimate of 
${\bf m}_{(lm)}$ and the uncertainty in each multipole is determined by 
${\bf N}_{(lm)(lm)^{\prime}}$.

The normal equations are ill-conditioned because ${\bf N}^{-1}$ has null 
vectors induced by the cut pixels. Fortunately, we do not need to compute 
${\bf m}$, but rather ${\bf z} \equiv ({\bf S} + {\bf N})^{-1} {\bf m}$.  
We can do this by solving the linear system
\begin{equation}
({\bf S} + {\bf N}){\bf z} = {\bf m} = {\bf N}{\bf y}
\label{zeq}
\end{equation}
We can put this into a computationally more tractable form by multiplying both
sides by ${\bf S}^{1/2}{\bf N}^{-1}$, giving
\begin{equation}
\left( {\bf I} + {\bf S}^{1/2}{\bf N}^{-1} {\bf S}^{1/2} \right) {\bf
S}^{1/2} {\bf z} = {\bf S}^{1/2} {\bf y}
\end{equation}
The reason for putting the system in this form is twofold.  First, it only 
involves ${\bf N}^{-1}$, not ${\bf N}$, which cannot be easily evaluated.  
Second, the matrix ${\bf A} \equiv {\bf I} + {\bf S}^{1/2}{\bf N}^{-1}
{\bf S}^{1/2}$ (which we will apply the conjugate gradient technique to) is
well conditioned in the sense that its eigenvalues only span a few orders of 
magnitude.  The second term -- which is explicitly the signal to noise ratio -- 
is regularized by the presence of the identity matrix.  Alternative forms 
such as $\left({\bf N}^{-1} + {\bf S}^{-1}\right){\bf S z} = {\bf y}$ do not 
share this property -- these matrices have eigenvalues that span a wide 
dynamic range. 

We obtain a good preconditioner by splitting the matrix into two parts
\begin{equation}
\widetilde{{\bf A}} = \left( 
\begin{tabular}{cc}
${\bf I}+{\bf S}^{1/2}\widetilde{{\bf N}}^{-1}{\bf S}^{1/2}$ & 0 \\ 
0 & ${\bf I} + {\rm diag}({\bf S}^{1/2}\widetilde{{\bf N}}^{-1}{\bf S}^{1/2})$
\end{tabular}
\right)
\label{preconditioner}
\end{equation}
where $\widetilde{{\bf N}}^{-1}$ is a sparse, approximate form of 
${\bf N}^{-1}$, given below.  The lower right hand block is used for large 
$l$ where the noise  dominates the signal, so that ${\bf S}^{1/2}
\widetilde{{\bf N}}^{-1} {\bf S}^{1/2}$ is small.  In this regime, we obtain 
${\bf z} \approx {\bf y}$, the spherical harmonic transform of the inverse 
variance weighted map.  At small $l$ the signal dominates, so we need to be 
able to approximate the form of the full noise matrix.  For {\it MAP} 2-year 
data, we find that the above preconditioner works well if we split the matrix 
at $l_{\rm cutoff} = 512$.  

We now make a brief detour to explore the structure of the inverse noise
matrix by expanding it in terms of the Wigner 3-$j$ symbols. This gives us
both a computationally efficient method to evaluate ${\bf N}^{-1}$, and, more
importantly, provides some insight into the structure of ${\bf N}^{-1}$ due to 
the sky cut and noise pattern. We first expand the inverse variance (weight) 
map in spherical harmonics. The multipole moments of this map are
\begin{equation}
{\bf w}_{lm}=\sum_i\frac{Y_{lm}^{*}(\Omega_i)}{\sigma_i^2}.
\label{noisealm}
\end{equation}
Using the completeness of the spherical harmonics, we may expand the inverse
noise matrix, equation (\ref{Noise}), as
\begin{eqnarray}
{\bf N}_{(lm)(lm)^{\prime}}^{-1} & = &
\sum_{(lm)^{\prime\prime}} {\bf w}_{(lm)^{\prime\prime}}
\sum_i Y_{lm}(\Omega_i) Y_{l^{\prime}m^{\prime}}(\Omega_i) 
Y_{l^{\prime\prime}m^{\prime \prime}}(\Omega_i) \label{clebsch} \\
 & = & \sum_{(lm)^{\prime\prime}} {\bf w}_{(lm)^{\prime\prime}}
\left( \frac{(2l+1)(2l^{\prime}+1)(2l^{\prime\prime}+1)}{4\pi}\right)^{1/2}
\left( 
\begin{tabular}{ccc}
$l$ & $l^{\prime}$ & $l^{\prime\prime}$ \\ 
0 & 0 & 0
\end{tabular}
\right) \left( 
\begin{tabular}{ccc}
$l$ & $l^{\prime}$ & $l^{\prime\prime}$ \\ 
$m$ & $m^{\prime}$ & $m^{\prime\prime}$%
\end{tabular}
\right)  \nonumber \label{Wigner}
\end{eqnarray}
where the terms in brackets are the 3-$j$ symbols. The first symbol is only
non-zero when $|l-l^{\prime}|\leq l^{\prime\prime}\leq |l+l^{\prime}|$.
The second symbol imposes the additional constraint that $m-m^{\prime}+
m^{\prime\prime}=0$. We use numerically stable recurrence relations
(Schulten, Klaus \& Gordon, 1975) to compute the symbols.  Alternatively, 
${\bf N}^{-1}$ may be computed by direct summation, which also requires 
$O(N^{3/2})$ operations.

This expansion suggests a simple approximation to the weight matrix. The 
dominant feature in the weight map is the galactic sky cut which is, to first
order, azimuthally symmetric in galactic coordinates.  In the limit of 
pure azimuthal symmetry the weight matrix would be block diagonal (proportional
to $\delta_{mm^{\prime}}$). This suggests a preconditioner of the form
\begin{equation}
\widetilde{{\bf N}}^{-1}_{(lm)(lm)^{\prime}} \equiv 
{\bf N}^{-1}_{(lm)(lm)^{\prime}} \delta_{mm^{\prime}}.
\end{equation}
Since $({\bf I}+{\bf S}^{1/2}\widetilde{{\bf N}}^{-1}{\bf S}^{1/2})$ is block
diagonal, we can compute its inverse in $O(N^2)$ steps. It is actually
significantly less than this, due to the decreasing size of the block 
matrices. The matrix $\widetilde{{\bf A}}^{-1}$ requires the largest amount 
of memory for storage, $O((N/20)^{3/2})$, where the savings in the prefactor 
result from the facts that 1) we only need the full matrix up to $l=512$, 2) 
all of the block matrices are symmetric, 3) the blocks are of decreasing size 
as $m$ increases, and 4) parity is preserved -- the covariance between even 
and odd $l$ terms vanishes. For a 2 million pixel map, only 100 MB of
memory is required to store the preconditioner in double precision.

\begin{figure}
\epsscale{1.00}
\plotone{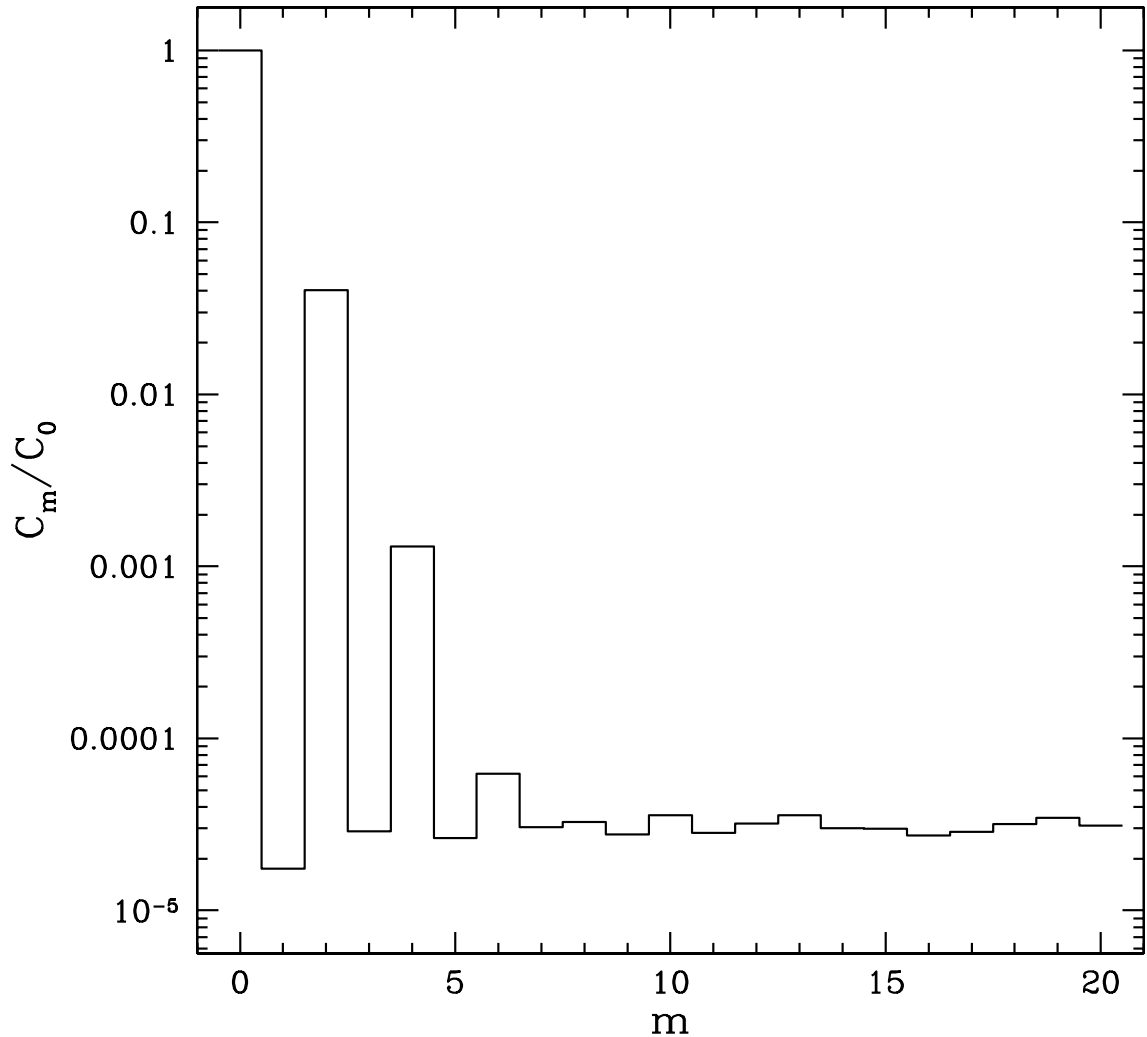}
\caption{Normalized $c_{m}$ vs $m$, where $c_{m} \equiv \sum_{l=m}^{l_{max}} 
|{\bf w}_{lm}|^2/(l_{max}+1-m)$ and ${\bf w}_{lm}$ is the spherical harmonic 
expansion of the weight map $1/\sigma_i^2$. Note that only the $m=0,2,4,6$ 
terms are significant, implying that ${\bf N}^{-1}$ is in fact very sparse. 
The baseline extending out to high m is contributed by the pixels cut due to
point sources; it disappears if there are no such cuts.} 
\label{noise_expansion}
\end{figure}

The block diagonal approximation is an excellent ansatz to which we need only 
apply small perturbative corrections.  Why does it work so well?  To answer 
this, we must understand the sparsity pattern of ${\bf N}^{-1}$ or, 
equivalently, its diagnostic ${\bf w}_{(lm)}$.  The {\it MAP} weight map is
approximately axisymmetric in ecliptic coordinates.  Rotation to galactic 
coordinates (which imposes a tilt of about $60^{\circ}$) introduces a smooth, azimuthal variation in the noise pattern that 
can be perturbatively expanded in a Fourier series $e^{im\phi}$. The 
approximation of galactic axisymmetry used in the preconditioner is the 
largest ($m=0$) term in such an expansion.  We can quantify the fall-off in 
${\bf w}_{(lm)}$ with increasing $m$ by computing the ``power'' at each $m$,
defined as
\begin{equation}
c_{m} \equiv \frac{1}{(l_{max}+1-m)} \sum_{l=m}^{l_{max}}|{\bf w}_{lm}|^{2}.
\end{equation}
Figure \ref{noise_expansion} shows that corrections to the $m=0$ mode are at 
most a few percent. The only high frequency contribution to the weight
map comes from point sources. Since they only occupy $\sim 5\%$ of the
sky, their effect is small.

A note on overall computational cost: each iteration of the conjugate gradient 
routine requires forming the products ${\bf A}{\bf w}_1$ and 
$\widetilde{{\bf A}}^{-1}{\bf w}_2$ for some work vectors ${\bf w}_1$ and 
${\bf w}_2$.  The former involves 2 products with diagonal matrices 
(${\bf S}^{1/2}$) and a forward and inverse spherical harmonic transform, 
each $O(N^{3/2})$.  The latter involves a matrix-vector multiplication where 
$\widetilde{{\bf A}}^{-1}$ has fewer than $O(N^{3/2})$ non-zero elements. Thus,
the cost per iteration is $O(N^{3/2})$, and the total cost of solving the 
linear system is $O(N_{iter}N^{3/2})$. A good preconditioner is the key to 
minimizing $N_{iter}$ -- in its absence, $N_{iter} \sim N$.  We find that the
linear system, equation (\ref{zeq}), can be solved in about 6 iterations for 
the specifications appropriate to the 2-year {\it MAP} data and the 
preconditioner specified above. This requires $\sim$30 seconds to solve for a 
500,000 pixel map, and $\sim$ 250 seconds to solve for a 2 million pixel map, 
running as a single processor job on an SGI Origin 2000.

Note that the preconditioner we have described is optimized for the
MAP experiment. The choice of the appropriate preconditioner is likely
to vary from experiment to experiment, in particular due to the
different weight matrix ${\bf N}^{-1}$ of each experiment, depending on
its survey geometry and noise properties. Since the normal
equations for the time-ordered data are solved in the map-making
process, the weight-matrix ${\bf N}^{-1}$ already arises naturally in
the previous step of the pipeline (Wright 1996). Its sparsity
properties are therefore fairly well understood already at this point,
and may be exploited in the construction of the preconditioner. This
is another advantage of our approach, compared to previous techniques
cast in terms of ${\bf N}$. Note that our use of fast transforms to
reduce the cost of matrix multiplies to $O(N^{3/2})$ is fairly
general. In the complete absence of any intuition about symmetries in
the noise and geometry of the experiment, one can still use the fact
that almost all the matrices we deal with in CMB experiments are
diagonally dominant (the correlation function of both the signal and
noise fall rapidly with separation in any experiment, and thus
off-diagonal elements die rapidly). Thus, one possibility is to approximate a matrix by its diagonal. More general techniques for sparse symmetric positive definite matrices exist,
e.g. Cholesky multifrontal methods (Liu 1989), which can be used to build preconditioners.        

\section{Computation of the Trace}

While we can use the preconditioner to compute ${\bf m}^T{\bf C}^{-1}
{\bf P}^l{\bf C}^{-1}{\bf m}$ rapidly, it is still very time consuming to 
evaluate ${\rm tr}({\bf C}^{-1}{\bf P}^l)$ for each value of $l$ by 
iteratively solving a linear equation for each $(lm)$ term in the trace. There 
are two approaches to avoiding a brute force evaluation of the full trace. 1) Approximate 
${\rm tr}({\bf C}^{-1}{\bf P}^l)$ as ${\rm tr}({\bf S}^{-1/2}
\widetilde{{\bf A}}^{-1}{\bf S}^{1/2}\widetilde{{\bf N}}^{-1}{\bf P}^l )$
using the fact that ${\bf C}^{-1} = {\bf S}^{-1/2}{\bf A}^{-1}{\bf S}^{1/2}
{\bf N}^{-1}$. This returns a high quality estimate of $c_l$. 2) Evaluate
the trace for a given $c_l$ using Monte Carlo simulations of maps drawn 
from that spectrum.  This approach exploits the fact that
\begin{equation}
{\rm tr}({\bf C}^{-1}{\bf P}^l) = \left\langle {\bf m}^T{\bf C}^{-1}{\bf P}^l
{\bf C}^{-1}{\bf m}\right\rangle
\end{equation}
where we have used $\langle {\bf m}{\bf m}^T\rangle = {\bf C}$. This is the 
most computationally expensive part of our algorithm: it requires 
$O(N_{mc}N_{iter}N^{3/2})$ operations.  Therefore we first maximize the 
likelihood function using the approximate trace (method 1), only switching to 
the Monte Carlo evaluation of the trace once the former solution has converged. 
Since we are very close to the true answer at this point, the Monte Carlo
solution converges very quickly. Note that the Monte Carlo method is 
guaranteed by construction to be unbiased. Additionally, we can include 
non-linear effects in the synthetic maps that are not easily modeled in the 
covariance matrix ${\bf C}$. This will both correctly alter the maximum 
likelihood point and propagate through to the error estimates. We can further 
generalize the method to incorporate all known effects in our analysis by 
simulating the full analysis pipeline, rather than just the processed map.

How many Monte Carlo evaluations of ${\bf q}^{(i)} \equiv {\bf m}^{(i)}
{\bf C}^{-1}{\bf P}^l{\bf C}^{-1}{\bf m}^{(i)}$, where $i$ is a realization
index, are necessary to obtain an accurate determination of ${\rm tr}
({\bf C}^{-1}{\bf P}^l)$?  To answer this question, we need to determine how 
the variance in ${\bf q}^{(i)}$ propagates through to variance in the recovered 
$c_l$. Suppose we were able to calculate $\langle{\bf q}\rangle$ exactly for 
a given power spectrum $c_l$ (e.g., by using an infinite number of simulations). 
The variance in our recovered $c_l$ would then be given by the Cramer-Rao 
minimum variance bound
\begin{eqnarray}
\left\langle ({\bf c}^{\rm recovered}-{\bf c}^{\rm true})
({\bf c}^{\rm recovered}-{\bf c}^{\rm true})^T\right\rangle & = & 
\frac{1}{4}{\bf F}^{-1} \left\langle ({\bf q}-\langle{\bf q}\rangle)
({\bf q}-\langle{\bf q}\rangle)^T\right\rangle {\bf F}^{-1}  \nonumber \\
& = & {\bf F}^{-1}
\end{eqnarray}
which implies 
\begin{equation}
\left\langle ({\bf q}-\langle{\bf q}\rangle)({\bf q}-\langle{\bf q}\rangle)^T
\right\rangle 
= 4{\bf F}
\end{equation}

Now we do not know $\langle{\bf q}\rangle$, but $\langle{\bf q}\rangle^{\prime}$, 
obtained by averaging ${\bf q}^{(i)}$ over ${\it N_{mc}}$ uncorrelated Monte 
Carlo simulations. Hence, $\langle{\bf q}\rangle^{\prime}$ has a Gaussian 
distribution, with variance
\begin{equation}
\sigma^2_{\langle{\bf q}\rangle^{\prime}} 
= \frac{1}{N_{mc}}\sigma^2_{{\bf q}^{(i)}}
\end{equation}
which implies 
\begin{equation}
\left\langle({\bf c}^{\rm recovered}-{\bf c}^{\rm true}) 
({\bf c}^{\rm recovered}-{\bf c}^{\rm true})^T\right\rangle = 
\left(1+\frac{1}{N_{mc}}\right) {\bf F}^{-1}
\end{equation}
where we have added the errors in quadrature. Thus, if we average 100 Monte Carlos to evaluate $\langle{\bf q}
\rangle^{\prime}$, our errors will be only 0.5\% larger than the minimum
variance case.

\begin{figure}
\epsscale{1.00}
\plotone{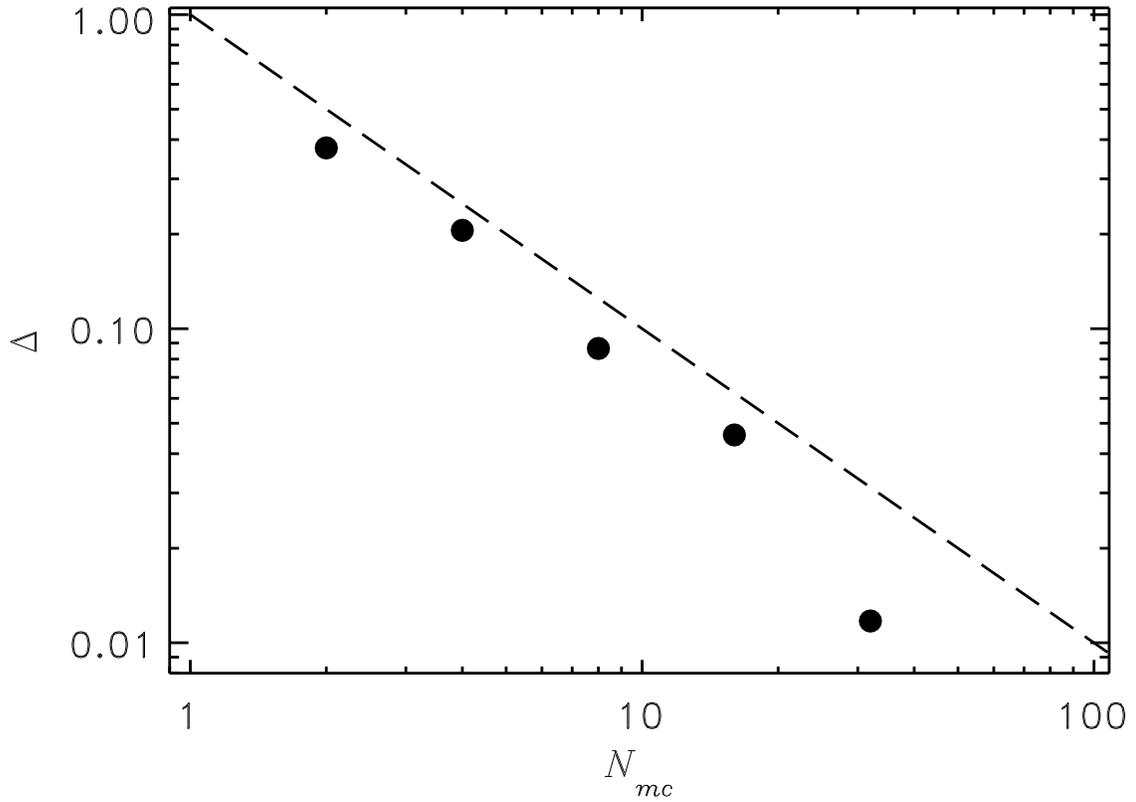}
\caption{Plot of $\Delta \equiv \chi^{2}/(l_{max}-1)-1$ vs. number of Monte 
Carlo simulations used to compute ${\rm tr}({\bf C}^{-1} {\bf P}^{l})$.  This 
result is based on solving for $c_{l}$ from a given $512\times 1024$ pixel map. 
Our expectation that $\Delta \propto 1/N_{mc}$ (dashed line) is met.}
\label{chisq}
\end{figure}

We have verified these expectations numerically. For a $512 \times 1024$ pixel 
test map (note that this is smaller than the $1024 \times 2048$ map we analyze 
in subsequent sections), we recover $c_l$ by maximizing the likelihood 
function, but in each instance we compute $\langle {\bf q}\rangle ^{\prime }$ 
using a different number of Monte Carlo simulations. For each solution 
$c_l^{(N_{mc})}$ we then compute
\begin{equation}
\chi_{N_{mc}}^2 = ({\bf c}^{(N_{mc})}-{\bf c}^{\rm true})^T{\bf F}
({\bf c}^{(N_{mc})}-{\bf c}^{\rm true})
\end{equation}
which is expected to have a mean of $(1 + 1/N_{mc})(l_{max}-1)$. For 
consistency we must use the same Fisher matrix for each $N_{mc}$, but this 
only affects the overall normalization of $\chi_{N_{mc}}^2$. Here we have 
used the Fisher matrix of the true underlying power spectrum. Figure 
\ref{chisq} shows that to a very good approximation
\begin{equation}
\left( \frac{\chi^2_{N_{mc}}}{l_{max}-1}\right) \propto 1+\frac{1}{N_{mc}}.
\end{equation}
Note that the quadratic convergence of the Newton-Raphson method means
that the number of significant digits is doubled at each iteration. We
can match this rate of convergence by doubling the number of Monte Carlo 
simulations used to compute the trace at each iteration. This saves some 
computer time in the early iterations when high precision is not required.

\section{The Fisher Matrix and Error Estimates}

As with the trace, a good approximation to the true Fisher matrix is given by
\begin{equation}
\widetilde{\bf F}_{ll^{\prime}} = \frac{1}{2}{\rm tr}\left(
\widetilde{\bf C}^{-1}{\bf P}^l\widetilde{\bf C}^{-1}{\bf P}^{l^{\prime}}\right) 
\label{fisherapprox} 
\end{equation}
where $\widetilde{\bf C}^{-1} = {\bf S}^{-1/2}\widetilde{{\bf A}}^{-1}
{\bf S}^{1/2} \widetilde{{\bf N}}^{-1}$.  Since $\widetilde{\bf A}^{-1}$ and 
$\widetilde{\bf N}^{-1}$ have already been computed when forming the 
preconditioner, and since ${\bf P}^{l}$ consists only of zeros and ones along 
the diagonal, calculating this approximate Fisher matrix only entails the 
multiplication of block diagonal matrices, which is relatively quick.  It is
important to note that the Newton-Raphson method only requires an approximate 
Fisher matrix to converge to the maximum likelihood solution -- the final 
solution is independent of the Fisher matrix.

Of course any subsequent analysis of the power spectrum, such as cosmological
parameter fitting, requires accurate error estimates for the $c_l$.  These are
often obtained by assuming the $c_l$ are Gaussian distributed and computing 
their covariance matrix ${\bf F}^{\rm -1}$.  This description is likely to be
fine for large $l$ (say $l > 32$) but for small $l$ we would like a better
description. This is because the number of independent modes (which
scales as $2l+1$) is small
for low $l$, and only at large $l$ does the Central Limit Theorem
guarantee that the $c_{l}$'s are Gaussian distributed. One could attempt to compute the distribution of the $c_l$ 
directly from the Monte Carlo simulations, {\it independent} of any 
assumptions of Gaussianity. Specifically, for low $l$ we should work with
smoothed maps (with a small number of pixels), which may be solved in
a matter of seconds. For each realization $i$, one can find the maximum
likelihood power spectrum $c_l^{(i)}$ by the same procedure used to
solve for the full map. By making thousands of realisations, one
immediately obtains the full probability distribution of the $c_{l}$'s,
including higher order moments, and not just the second moment (as in
the Fisher matrix formalism). If we
sort the recovered $c_l^{(i)}$'s, we immediately obtain (in general asymmetric)
confidence intervals. This Monte-Carlo method of computing the
distribution of errors makes no assumptions about the shape of the
likelihood function, apart from having an identifiable maximum. Note
that each realisation requires fresh computations of ${\rm tr}({\bf
C}^{-1} {\bf P}^{l})$ in order to iterate to convergence. It is thus
prohibitively expensive (and unnecessary) for high $l$, and we
recommend this procedure up to $l=32$. 

\begin{figure}
\epsscale{1.00}
\plotone{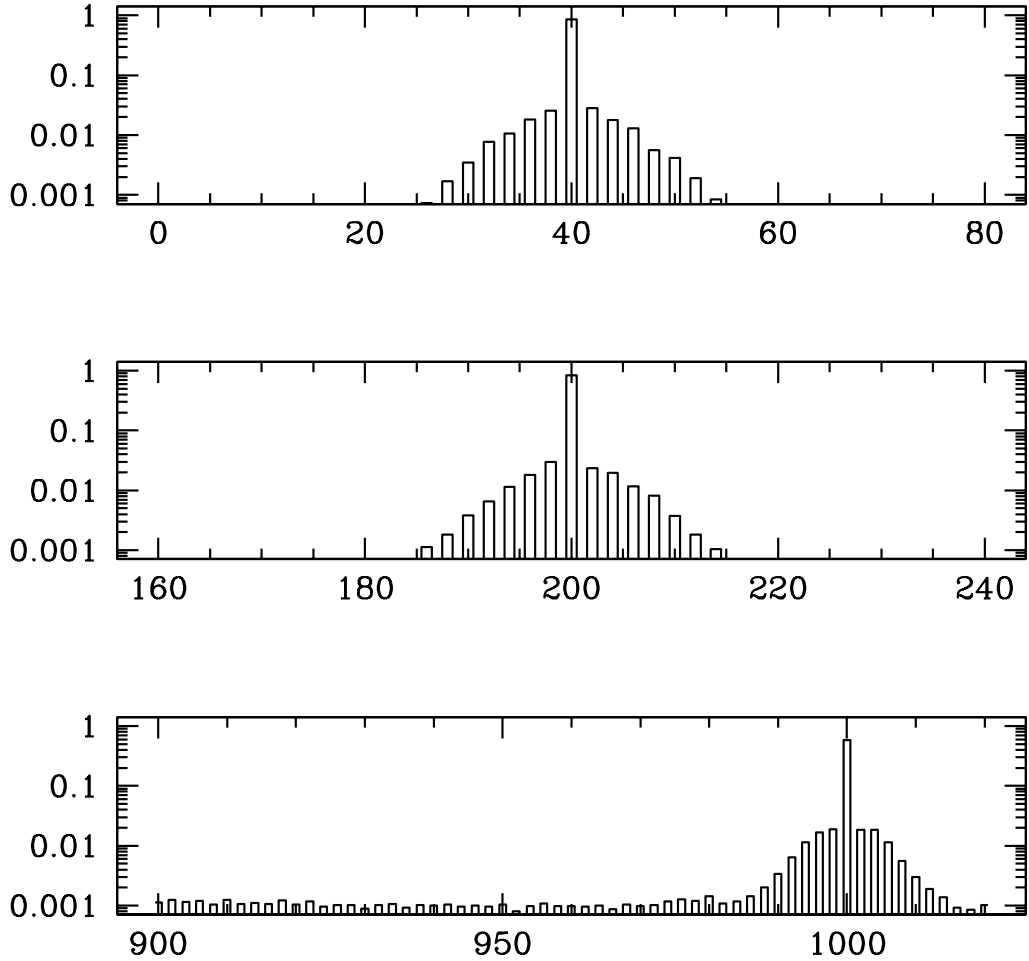}
\caption{Window functions at various $l$. At high signal to noise, we observe 
the well-known translational independence of the window function. The small 
size of the galactic cut allows for an extremely narrow window function. At 
low signal to noise correlations between widely separated $l$ appear, but 
remain very weak.}
\label{window}
\end{figure}

For higher order multipoles, we need only compute the Fisher matrix. In 
principle we can use our Monte Carlo results to compute
\begin{equation}
{\bf F}_{l l^{\prime}}=\frac{1}{2} \left\langle{\bf m}^T{\bf C}^{-1}{\bf P}^l
{\bf C}^{-1}{\bf P}^{l^{\prime}}{\bf C}^{-1}{\bf m}\right\rangle.
\label{NaiveFisher}
\end{equation}
Unfortunately this is a factor of $N_{iter}l_{max}$ times more expensive than
computing ${\rm tr}({\bf C}^{-1}{\bf P}^l)$, which is prohibitive. However, we 
have already noted that
\begin{equation}
{\bf F} = \frac{1}{4} \left\langle({\bf q} - \langle{\bf q}\rangle)
({\bf q} - \langle{\bf q}\rangle)^T \right\rangle
\end{equation}
Thus, in principle, one can obtain the Fisher matrix from the Monte Carlo 
simulations for free. In practice, we have found this to be a good recipe 
for computing the (comparatively large) diagonal elements, but to be too noisy 
for use in computing the (comparatively small) off-diagonal elements. To 
overcome this, we use the fact that the shape of the Fisher matrix (which also yields the 
window function for the $c_l$) depends mainly on the geometry of the weight map 
and only weakly on other aspects of the map. We therefore extract the shape of 
the approximate Fisher matrix $\widetilde{\bf F}$ from equation 
(\ref{fisherapprox}) and renormalize to the variances obtained from the Monte 
Carlo simulations ${\bf F}^{\rm MC}$ i.e.
\begin{equation}
{\bf F}_{l l^{\prime}} = \widetilde{\bf F}_{l l^{\prime}} \left(
\frac{{\bf F}^{\rm MC}_{l l}{\bf F}^{\rm MC}_{l^{\prime} l^{\prime}}}
{\widetilde{\bf F}_{l l}\widetilde{\bf F}_{l^{\prime} l^{\prime}}} \right)^{1/2}
\label{Fisher_rescaled}
\end{equation}
Note that for large $l$, where the signal-to-noise becomes low due to beam
smearing, the off-diagonal terms in the Fisher matrix start to become
significant.  This is easy to understand intuitively: the statistics of the 
$c_{l}$ at large $l$ are dominated by noise, which is not rotationally 
invariant (see Appendix A). This induces correlations amongst the multipoles. 
However, for the signal-to-noise ratio of the 2-year {\it MAP} data, this 
effect is unimportant (see Figure \ref{window}). We find that our results
for $\chi^{2}$ change by very little if we approximate the Fisher matrix as 
diagonal.

We emphasize that our {\it final} Fisher matrix is extremely accurate
and may be used in good faith for parameter estimation. The calculation of 
the diagonal elements is exact and may be performed to arbitrary accuracy 
by increasing the number of Monte Carlo simulations. The scaling of the 
off-diagonal elements is approximate and uses the azimuthally averaged noise 
map. However, the off-diagonal elements make a small contribution to the 
Fisher matrix to begin with, as quantified by the small change in $\chi^{2}$. 
The effect of excluding non-azimuthal corrections, which are down by another 
two orders of magnitude (Figure \ref{noise_expansion}), is therefore utterly 
negligible. Thus, while it is possible to make perturbative corrections to 
the scaling of the off-diagonal elements, this is unnecessary in practice. 
Another way of saying this is that the dominant source of cross talk among 
multipoles in the Fisher matrix is the galactic cut. Note that the neglect of 
azimuthal noise variations would {\it not} be valid for the diagonal elements
since it would lead to an underestimate of the variance. 

One should be careful in the choice of $c_{l}$ used to compute the final 
Fisher matrix. It is ${\it not}$ correct to use the recovered $c_{l}$ 
because they are noisy. Any $l$ for which we 
recovered a low $c_l$ would be assigned a spuriously small cosmic
variance contribution to its error, and hence 
be given more weight in any subsequent analysis. This would consistently 
bias any fit to the power spectrum in the direction of less
power. This effect has been discussed by Bond, Jaffe \& Knox (1998), and Seljak (1997). We 
therefore invoke our prior expectation that the underlying power spectrum is 
smooth, and smooth the recovered $c_l$ with a spline prior to forming 
our final error estimates (see Appendix B for the smoothing
procedure).
                                                                                                                                      
\section{Estimating Cosmological Parameters}

If we have a cosmological model that appears to be a good fit to the data,
we can use the likelihood approach to directly compute cosmological 
parameters from a map. Rather than solve for the power spectrum $c_l$, we can
solve for a set of parameters, ${\bf p}$ = ($\Omega_b$, $\Omega_{CDM}$, $H_0$, 
etc.), that predict the spectrum.

Proceeding as in \S 2, we can maximize the likelihood function for ${\bf p}$ 
by expanding it as a Taylor series around its maximum at $\bar{\bf p}$
\begin{equation}
f \equiv -2 \ln {\cal L} = \bar{f} 
  + \sum_l \left.\frac{\partial f}{\partial p_k}\right|_{\bar{{\bf p}}}
    \left(p_k-\bar{p}_k\right) 
  + \sum_{k,k^{\prime}}\frac{1}{2}
    \left.\frac{\partial^2 f}{\partial p_k \partial p_{k^{\prime}}}
    \right|_{\bar{{\bf p}}}
    \left(p_k-\bar{p}_k\right)\left(p_{k^{\prime}}-\bar{p}_{k^{\prime}}\right)
\label{parameterTaylor}
\end{equation}
which leads to the analog of equation (\ref{Newton})
\begin{equation}
p_k^{(n+1)} = p_k^{(n)} 
 - \frac{1}{2}\sum_{k^{\prime}}{\bf F}^{-1}_{kk^{\prime}}
 \left.\frac{\partial f}{\partial p_{k^{\prime}}}\right|_{{\bf p}^{(n)}}.
\label{parameterNR}
\end{equation}
The derivatives of the likelihood function may be computed using the chain rule
\begin{equation}
\left.\frac{\partial f}{\partial p_k}\right|_{\bf p} 
= \sum_l \left.\frac{\partial f}{\partial c_l}\right|_{c_l({\bf p})}
  \left.\frac{\partial c_l}{\partial p_k}\right|_{\bf p}.
\end{equation}
Similarly, the Fisher matrix of the parameters is given by
\begin{eqnarray}
{\bf F}_{k k^{\prime}} & = & 
\left\langle -\left(\frac{\partial^2}{\partial p_k \partial p_{k^{\prime}}}\right) 
\ln{\cal L}\right\rangle \\
 & = & \sum_{ll^{\prime}}\frac{\partial c_l}{\partial p_k} 
\left\langle -\left(\frac{\partial^2}{\partial c_l \partial c_{l^{\prime}}}\right) 
\ln{\cal L}\right\rangle \frac{\partial c_{l^{\prime}}}{\partial p_{k^{\prime}}}
\nonumber \\
 & = & \sum_{ll^{\prime}} \frac{\partial c_l}{\partial p_k} {\bf F}_{l l^{\prime}} 
       \frac{\partial c_{l^{\prime}}}{\partial p_{k^{\prime}}}. \nonumber
\label{parameterFisher}
\end{eqnarray}
The model power spectra may be computed with a fast numerical code (Seljak \& 
Zaldarriaga 1996) and the partial derivatives ${\partial c_l}/{\partial p_k}$ 
may be computed using a finite difference approximation, typically using 
differences of order 2\% of each parameter's value. As before, one computes 
the first and second derivatives of the likelihood function with respect to 
the $c_{l}$ (equations (\ref{Gradlike}) and (\ref{Fisher})) via Monte Carlo 
techniques. 

In practice, using Newton-Raphson as a root finding technique in parameter 
space is less straightforward than when solving for the power spectrum $c_{l}$. 
The radius of convergence, the region in which the Taylor expansion is valid, 
is substantially smaller. Its scale is set by $({\bf F}^{-1})_{kk}^{1/2}$. In 
addition, the near degeneracies between various parameters create narrow 
valleys in likelihood space for which a Newton approach is not optimal. 
Techniques exist which could overcome these difficulties, e.g. the 
Levenberg-Marquardt method (Press et al. 1992), which employs a control 
parameter $\lambda$ to smoothly modulate between a steepest descent method 
($\lambda \gg 1$) and a Newton method ($\lambda \ll 1$)
\begin{equation}
F_{ll^{\prime}} = F_{ll^{\prime}}(1 + \lambda \delta_{ll^{\prime}}).
\end{equation}
Rather than developing such tools in this paper, we instead explore a simple 
$\chi^{2}$ fit to the power spectrum
\begin{equation}
\chi^{2}({\bf p}) = \sum_{ll^{\prime}}(c_l({\bf p})-c_l^{\rm recovered})
{\bf F}_{ll^{\prime}}(c_{l^{\prime}}({\bf p})-c_{l^{\prime}}^{\rm recovered}).
\end{equation}
We find that this gives an excellent approximation to maximizing the full
likelihood function from the map. To see why, consider the probability 
distribution of cosmological parameters given the recovered $c_{l}$
\begin{equation}
{\cal L}({\bf p}|c_l) \propto \frac{1}{\det{\bf F}^{-1}}
\exp\left[-\frac{1}{2}\delta {\bf c}({\bf p})^{T} \, {\bf F} \, 
\delta {\bf c}({\bf p})\right].
\end{equation}
$\chi^{2}$ fitting is a maximum likelihood estimator in the limit where 
${\bf F}$ is constant. Since ${\bf F}$ has contributions from the assumed 
cosmological signal (which is being varied) as well as from pixel noise, this 
is not strictly the case. However, as we demonstrate below, we determine the 
power spectrum with excellent precision at low $l$. For $l > 600$, where the 
confidence bands begin to broaden, the signal to noise drops rapidly due to 
beam smearing. In this regime fixed pixel noise is the dominant contribution 
to the variance of the $c_l$.  Overall, we have excellent knowledge of the 
Fisher matrix before we begin a $\chi^2$ fit, so holding ${\bf F}$ constant 
is an excellent approximation. A consistency check may be performed by 
comparing the power spectrum of the best fit cosmological model to the 
spline fit of the recovered power spectrum. If the two are very close, the Fisher matrices computed from 
either one will be virtually indistinguishable.

\section{Results}

In this section, we apply our numerical techniques to a realistic simulation
of the {\it MAP} data. We simulate two years of W band (94 GHz) data using
our current best estimate for the detector noise. We apply a Galactic plane 
cut that excludes the region $|b|<10^{\circ}$, and we cut an additional 5\% 
of the sky at random to simulate the effect of excising extragalactic point
sources. We assume that the noise is azimuthally symmetric in ecliptic
coordinates and that its variance changes by a factor of two between the
ecliptic pole and ecliptic plane. The maps are generated on a grid of 
galactic longitude and latitude points $(l,b)$ which result in smaller map 
pixels near the galactic poles (see Appendix A). We scale the noise per pixel 
inversely with pixel area, in addition to the intrinsic coverage variations.

\begin{figure}
\epsscale{1.00}
\plotone{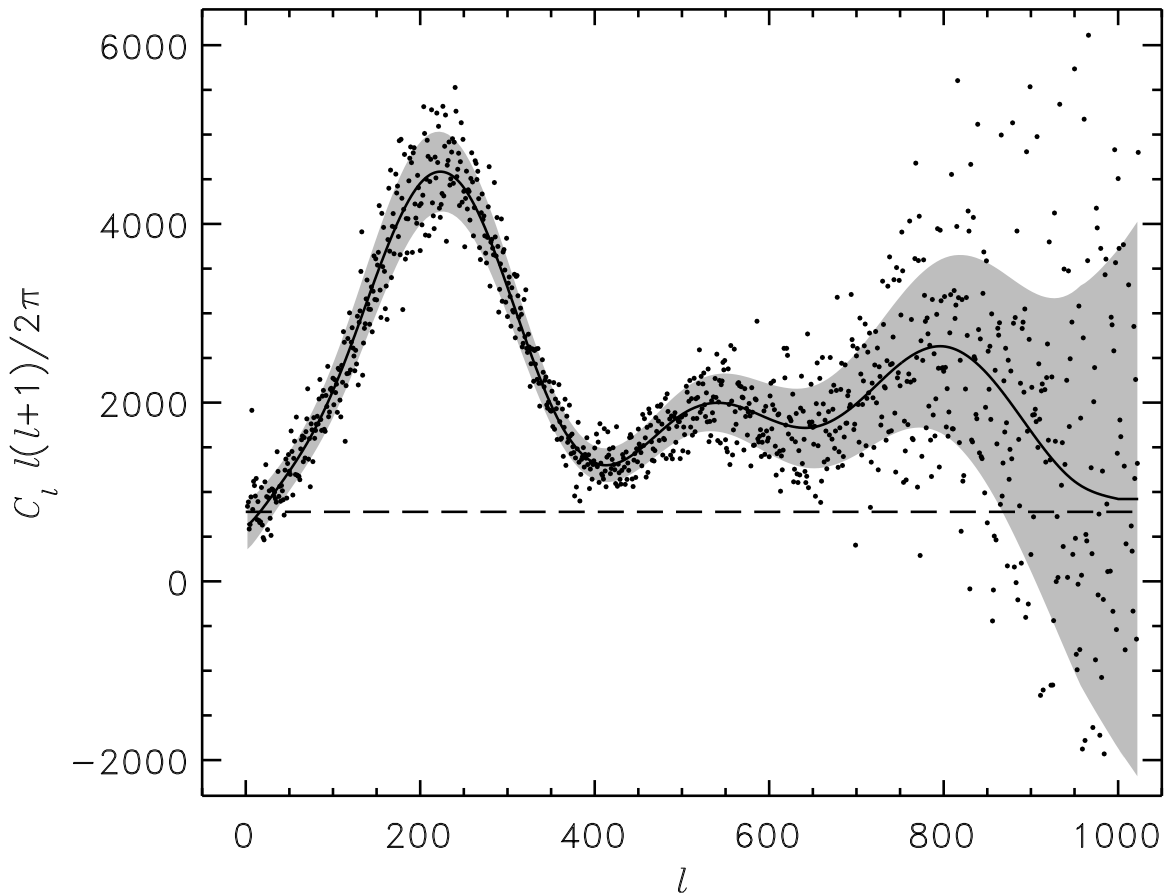}
\caption{Recovered power spectrum $c_{l}$ for a single simulated 
$1024\times2048$ pixel map. 10 Monte Carlo simulations were used to compute 
${\rm tr}({\bf C}^{-1} {\bf P}^{l})$. The grey band indicates the one sigma
uncertainty given by $({\bf F}_{ll}^{-1})^{1/2}$. The light dashed line 
indicates the starting guess. This run converged in 3 iterations using the 
approximate method and 2 subsequent iterations using the Monte Carlo method 
and took approximately 10 hours as an 8 processor job on an SGI Origin 2000 
computer.  See Figure \ref{spline} for an indication of how well the power
spectrum can be recovered under the assumption that it is a smooth function of
$l$.}
\label{main}
\end{figure}

We have obtained results for a 1024 by 2048 pixel map with power up to $l_{max}
= 1024$, using 10 Monte Carlo simulations to compute ${\rm tr}({\bf C}^{-1} 
{\bf P}^{l})$ and ${\bf F}$. With 10 simulations, our expected errors are 5\% 
larger than the minimum variance limit. The entire process converged in 5 
iterations and took 10 hours of cpu time on an 8 processor SGI Origin 2000 
computer. The recovered power spectrum is shown in Figure \ref{main}.  Note 
that some negative $c_{l}$ at high $l$ are to be expected. They reflect the 
fact that the variance in those modes was less than that expected from the 
noise alone.  One could adopt a prior distribution to prevent the $c_{l}$ 
from going negative. However, this would complicate the error analysis as 
the probability distribution of the $c_{l}$ would become skewed.
 
\begin{figure}
\epsscale{1.00}
\plotone{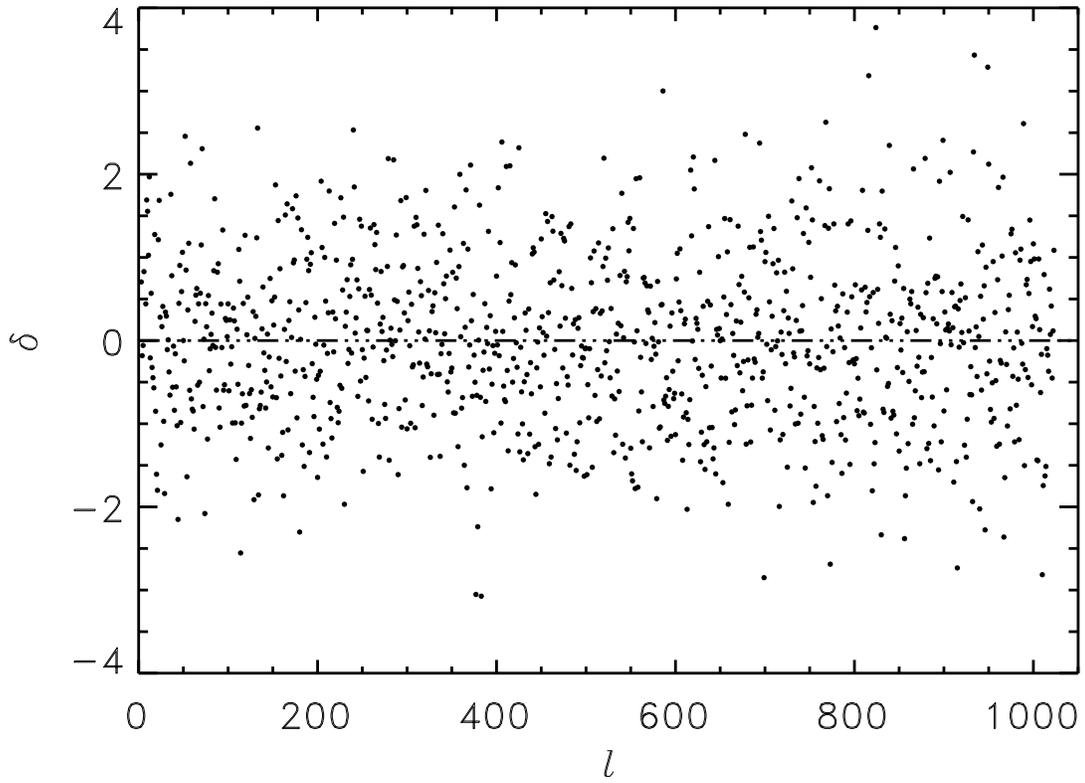}
\caption{Scatter plot of $\delta \equiv (c_l^{\rm recovered}-c_l^{\rm true})/
\sigma_l$, where $c_l^{\rm true}$ is the input power spectrum, 
$c_{l}^{\rm recovered}$ is the recovered spectrum shown in Figure \ref{main},
and $\sigma_l \equiv ({\bf F}^{-1})_{ll}^{1/2}$.}
\label{scatter}
\end{figure}

\begin{figure}
\epsscale{1.00}
\plotone{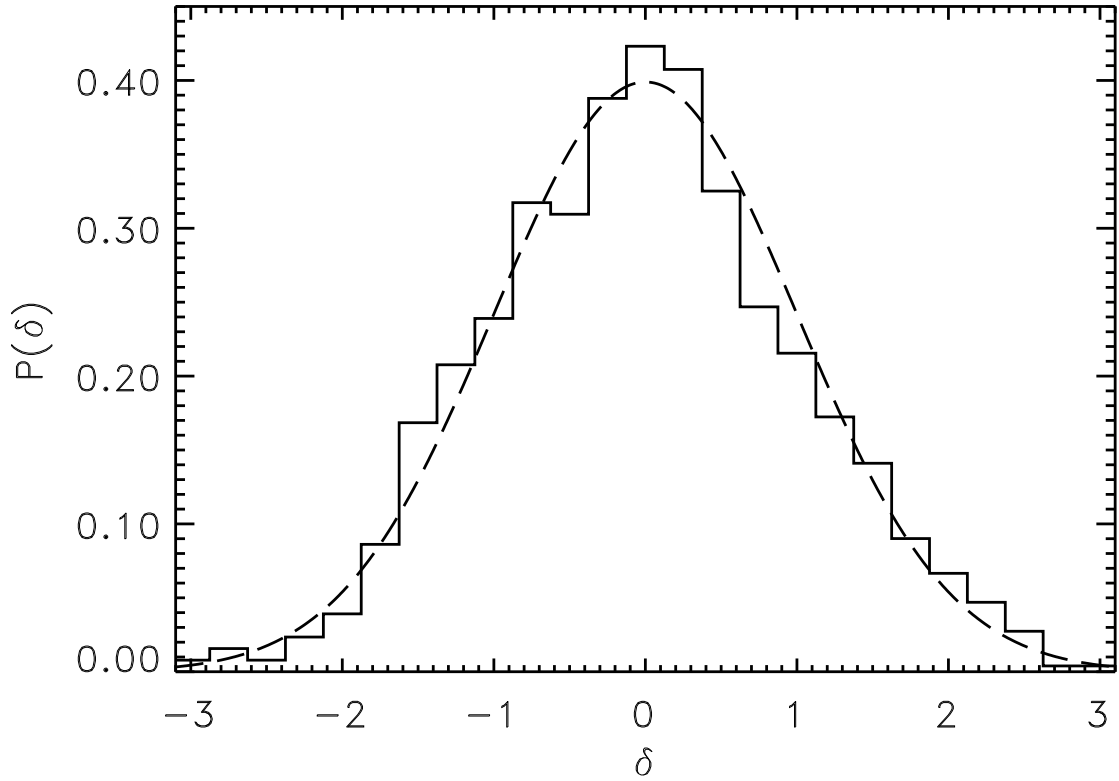}
\caption{Histogram plot of the distribution of $\delta \equiv
(c_l^{\rm recovered}-c_l^{\rm true})/\sigma_l$. Also depicted is a Gaussian 
distribution of zero mean and unit variance; the two distributions agree
closely. 67\% of the points lie between $\delta=-1$ and $\delta=+1$, and 95\% 
of the points lie between $\delta=-2$ and $\delta=+2$.}
\label{histogram}
\end{figure}

Since we know the true input power spectrum, we can quantify the goodness of 
our fit by computing
\begin{equation}
\chi^2 = \sum_{l,l^{\prime}} \delta c_l {\bf F}_{ll^{\prime}} \delta c_{l^{\prime}}
\end{equation}
where $\delta c_l \equiv c_l^{\rm recovered} - c_l^{\rm true}$, and ${\bf F}$
is the Fisher matrix of the input spectrum $c_l^{\rm true}$. We compute 
${\bf F}$ using equation (\ref{Fisher_rescaled}), with 128 Monte Carlo
simulations to compute ${\bf F}_{ll}$. We use a large number of Monte Carlo 
simulations to ensure that we are comparing our results to the true minimum 
variance Fisher matrix. With $N_{dof} \equiv l_{max}-1 = 1023$, we obtain 
$\chi^2/N_{dof} = 1.08$, in accordance with our expectation that it lie within 
the range $(1 + 1/N_{mc})(1 \pm \sqrt{2/N_{dof}}) = 1.1 \pm 0.05$ (1 $\sigma$).
We find that 67\% of the points lie within the 1 $\sigma$ error band, and 
95\% lie within the 2 $\sigma$ band.  In Figures \ref{scatter} and 
\ref{histogram} we plot $\delta \equiv \delta c_l / \sigma_l$, where 
$\sigma_l^2 \equiv ({\bf F}^{-1})_{ll}$ is the variance of each $c_l$. The 
distribution of errors is evidently Gaussian and has a mean value 
$\langle\delta\rangle=0.01$ and a standard deviation $\langle(\delta -
\langle\delta\rangle)^2\rangle=1.04$, indicating that our results are both 
unbiased and minimum variance.

A visually more impressive way to display the power spectrum recovery is to 
fit a smoothing spline to the recovered points.  The details of the spline fit
are given in Appendix B and the results are shown in Figure \ref{spline}.
Note that exploiting the prior that $c_{l}$ is a smooth function of $l$ allows 
us to come spectacularly close to the form of the underlying power spectrum. 
As described in Appendix B, we generate confidence regions on the fit by 
fitting splines to 128 Monte Carlo simulations of $c_l$ and sorting the fits 
at each $l$. The spline smoothing parameter has been chosen objectively 
using a process called cross-validation, which is a bootstrap technique 
(see Appendix B). If different criteria are used to compute the smoothing 
parameter, wiggles may appear in the fit (generally at high $l$), but it 
will generally stay within the depicted confidence bands.

\begin{figure}
\epsscale{1.00}
\plotone{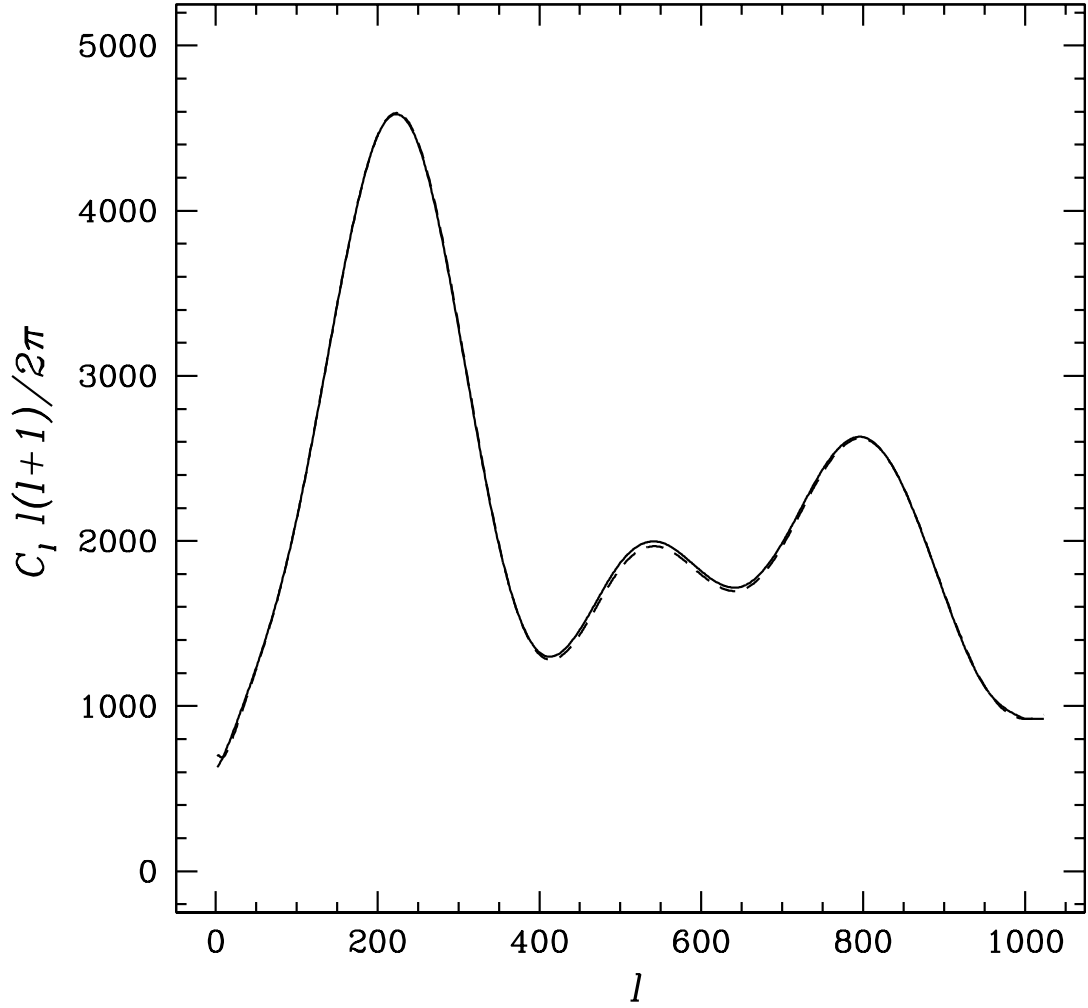}
\caption{The solid line indicates the input power spectrum, while the dashed 
line indicates the best-fit cosmological model from the data. The two are 
virtually identical.}
\label{paramcl}
\end{figure}

To estimate cosmological parameters, ${\bf p}$, we can use the spline fit as 
our best guess power spectrum for computing the Fisher matrix ${\bf F}$.
(Since the spline fit power spectrum and the input power spectrum are 
virtually identical, we have simply reused the Fisher matrix of the input 
spectrum we have already computed.) We then minimize
\begin{equation}
\chi^2 = \sum_{ll^{\prime}} (c_l({\bf p})-c_l^{\rm recovered}) 
{\bf F}_{l l^{\prime}} (c_{l^{\prime}}({\bf p})-c_{l^{\prime}}^{\rm recovered})
\end{equation}
where $c_l({\bf p})$ is computed for a given parameter set ${\bf p}$ using 
CMBFAST (Seljak \& Zaldarriaga 1996). We consider 6 free parameters: 
$\Omega_b$, $\Omega_{CDM}$, $h$, $\tau$, $n$, and the normalization, and we
allow for a cosmological constant term, $\Omega_{\Lambda}$, to enforce a flat
universe: $\Omega_b + \Omega_{CDM} + \Omega_{\Lambda} \equiv 1$. We minimize 
$\chi^{2}$ using an adaptive non-linear least-squares routine in the PORT 
optimization package (Gay 1990) after diagonalizing the Fisher matrix so that 
$\chi^{2}$ may be written as a sum of squares. The input model used was 
standard Cold Dark Matter (sCDM) with parameter values 0.1, 0.9, 0.5, 0.0, 
1.0, respectively.  A starting guess was obtained by finding the best-fit 
model to the spline fit by eye (in fact, the true underlying model was 
unknown to one of us at the time). The minimization routine recovered 
parameter values 0.09, 0.76, 0.53, 0.13, and 1.02, respectively. The standard 
errors on these parameters, as given by the parameter Fisher matrix, are 
$7 \times 10^{-3}$, 0.1, 0.02, 0.2, and 0.015, respectively. The best fit 
cosmological model yields $\chi^2/N_{dof} = 1.078$ with respect to the data, to
be compared to 1.082 for the input model with respect to the data, both within 
the 2 $\sigma$ range of expected variations for $\chi^2$. The input power 
spectrum and the best fit model are plotted in Figure \ref{paramcl}; they 
are virtually indistinguishable. 

Note that since the power spectrum of the best-fit model agrees closely with 
the spline fit, we are being self-consistent in holding ${\bf F}$ constant.  
If this did not occur, a computation that allowed ${\bf F}$ to vary might be 
called for. However, to the extent that the spline represents a non-parametric 
estimate of the underlying power spectrum, such lack of agreement might 
indicate deficiencies in the model, which may not have enough degrees of 
freedom. If one cannot obtain a reasonable value of $\chi^{2}$, even with
additional model parameters, then it would be reasonable to rule out that 
class of cosmological models. 

\section{Summary}

We have developed an unbiased, minimum variance estimator for the power 
spectrum of temperature fluctuations in CMB maps. We have used it to
estimate the power spectrum up to $l=1024$ from a 2 million pixel map that 
simulates single frequency {\it MAP} data (94 GHz) with realistic instrument 
noise and sky cuts.  In contrast with existing algorithms, which require 
$O(N^{3})$ operations, our algorithm is $O(N^{2})$, and thus can be run 
overnight on existing workstations, rather than running for months at 
national supercomputing facilities. We anticipate further improvements in 
performance with more aggressive optimization (and parallelization) of the 
code. 

In addition, we have estimated cosmological parameters from the recovered 
$c_l$ and find we can recover the input model with good precision. Thus, the 
pipeline from differential time-ordered data to cosmological parameters is now 
complete and well within present day computational capabilities.

Possibilities for future work include extensions to multi-frequency and 
polarization data, as well as the inclusion of Galactic foregrounds and 
systematic effects, such as striping, in our treatment of the data. The 
former only incur a linear increase of order a few in the operations
count. The latter must be dealt with by inclusion of the appropriate new
terms in the covariance matrix, as well as subsequent simulation of
the systematic effects in the Monte-Carlo trace computations.

\section{Acknowledgements}

We have benefited from discussions with Chuck Bennett, Peter Kostelec,
Robert Lupton, Bill Press, George Rybicki, Michael Strauss, Max Tegmark, 
Michael Vogeley, and Ned Wright.  SPO, DNS, and GH are supported by NASA's 
{\it MAP} Project.

\begin{appendix}

\section{Fast Spherical Harmonic Transforms}

The algorithm presented in this paper relies heavily on the fact that we have
available fast, $O(N^{3/2})$, forward and inverse spherical harmonic
transforms on the sphere, as opposed to standard $O(N^2)$ transforms. 
We use a formulation which employs FFTs in $\phi$, and thus requires that the
map pixels be distributed on rows of constant $\theta$. The scheme was first 
brought to the attention of the CMB community by Muciaccia, Natoli \&
Vittorio (1998). Alternative formulations exist (Dilts 1985) which also use 
FFTs in the $\theta$ direction. The most sophisticated implementation 
requires only $O(N(\log N)^2)$ operations for both transforms, by using fast 
Legendre transforms (Driscoll \& Healey 1994, Healy, Rockmore \& Moore, 1996). 
In practice, due to cache problems in the use of precomputed data, current
implementations run no faster than the naive $O(N^{3/2})$ algorithms, 
though work is presently underway to remove this barrier (Kostelec 1998).
For completeness, we summarize the transform method below.

{\bf Making maps} -- Given a set of $a_{lm}$'s, we wish to evaluate
\begin{equation}
\delta t(\theta_i,\phi_j) = \sum_{l=0}^{l_{max}}
\sum_{m=-l}^{l} a_{lm} Y_{lm}(\theta_i,\phi_j).
\end{equation}
Now, observing that we can interchange the order of summation
\begin{equation}
\sum_{l=0}^{l_{max}}\sum_{m=-l}^{l}
 \longleftrightarrow \sum_{m=-l_{max}}^{l_{max}}\sum_{l=|m|}^{l_{max}}
\end{equation}
we can write 
\begin{equation}
\delta t(\theta_i,\phi_j) = \sum_{m=-l_{max}}^{l_{max}}
 e^{im\phi_j} q_m(\theta_i)
\end{equation}
where 
\begin{equation}
q_m(\theta_i) = \sum_{l=|m|}^{l_{max}} a_{lm}\,
 \sqrt{\frac{2l+1}{4\pi}\frac{(l-m)!}{(l+m)!}}\,P^l_m(\cos\theta_i).
\end{equation}
The Legendre functions $P^{l}_{m}$ may be generated using standard recursion 
relations. The obvious motivation for writing the expansion in this form is 
to employ FFT techniques in the evaluation.

What is our total operations count?  At fixed $\theta_i$, we need to generate 
$l_{max}^2$ $q_{m}$'s. (Note therefore that our memory storage requirements
are $O(N)$, which is entirely feasible).  Since there are $n_{\theta}$ 
$\theta_i$'s to step through, the total operations count is 
$l_{max}^2 n_{\theta} \sim N^{3/2}$.  The cost of the FFT at fixed $\phi_j$ 
is only $O(n_{\phi}\log n_{\phi})$.  Since there are $n_{\phi}$ $\phi_j$'s to 
step through, the FFT's only cost $O(N \log N)$.  Thus, to leading order, map 
making is an $O(N^{3/2})$ process.

{\bf Inverting maps} -- The formal inverse transform is defined in terms of 
an integral 
\begin{equation}
a_{lm} = \int Y_{lm}(\Omega) \delta t(\Omega) d\Omega
\end{equation}
which may be evaluated with a cubature formula
\begin{equation}
a_{lm} = \sum_{i} Y_{lm}(\Omega_{i}) \delta t(\Omega_{i})w(\Omega_{i})
\label{cubature}
\end{equation}
where the integration weight $w(\Omega_{i})$ is essentially the solid angle 
of pixel $i$.  Assuming that our pixels are equally spaced in $\phi$ at a given
$\theta$, this may be cast in the form
\begin{equation}
q_m(\theta_i) = \sum_j e^{-im\phi_j} \delta t(\theta_i,\phi_j)
\end{equation}
with
\begin{equation}
a_{lm} = \sum_i w(\theta_i)\,q_m(\theta_i)
 \sqrt{\frac{2l+1}{4\pi}\frac{(l-m)!}{(l+m)!}}\,P^l_m(\cos\theta_i)
\end{equation}
where $w(\theta_i)$ is proportional to the pixel solid angle at $\theta_i$.

In practice, we actually wish to evaluate expressions of the form
\begin{equation} 
\widetilde a_{lm}= \sum_{i} Y_{lm}(\Omega_{i})f(\Omega_{i})
\label{noweights}
\end{equation}
where $f(\Omega)$ is some function on the sphere.  These terms arise in the 
normal equations, (\ref{Noise}) and (\ref{tlm}), where $f$ is an inverse 
variance weighted temperature map, and in the factorization of matrices into 
convolutions of a diagonal matrix with the spherical harmonics, as in equation 
(\ref{pixelsignal}). This expression is distinct from the formal inverse 
transform, but may be evaluated using the same FFT methods by substituting $f$
for $\delta t$ and omitting the integration weights.

The reader can easily verify that inverting maps is also an $O(N^{3/2})$ 
process requiring only $O(N)$ memory storage.  One can also exploit various 
symmetries, such as $q_{-m} = q^*_m$, and $P^l_m(\cos(\pi-\theta)) = 
(-1)^{l+m} P^l_m(\cos\theta)$ to further speed up the transforms. In practice, 
we find the lion's share of cpu time for both the forward and inverse 
transforms is spent in generating the $P^{l}_{m}$. Since we already demand 
$O(N^{3/2})$ storage for the preconditioner matrix, it is only a modest 
increase to store the $P^{l}_{m}$ in memory, which is also an $O(N^{3/2})$ 
requirement. (Note that only one hemisphere need be stored.) This generally 
leads to an order of magnitude decrease in cpu time: for example, the time 
required for each $l=512$ transform is reduced from 25 s to 3 s. At present, 
we find that each $l=1024$ transform takes about $25s$ as a single processor 
job on an SGI Origin 2000. 

We close this Appendix with some comments about the role of pixels in the
evaluation of inverse spherical harmonic transforms. The cubature formula 
provides a formal check on pixelization schemes by ensuring that pixelization 
errors are negligible, i.e. that there is no information loss in going from a 
continuous to a discrete field. Formally this implies specifying a grid 
$\Omega_{i}$ and integration weights $w(\Omega_{i}))$ such that the 
integration error
\begin{equation}
\epsilon = \sum_{i=1}^{N}w(\Omega_i)f(\Omega_i)- \int f(\Omega) d\Omega 
\end{equation}
vanishes when $f$ is a polynomial of degree $D$. In this paper we employ a 
spherical product Gauss cubature formula for integrating over the sphere 
(see, e.g., Stroud 1971). This scheme has the property of requiring the 
minimum number of pixels for a given polynomial degree of any pixelization 
scheme. The grid points in $\phi$ are given by the zeroes of the Fourier 
basis sines and cosines (simply the equiangular grid), while the grid points 
in $\theta$ are given by the zeroes of the Legendre polynomial basis.  
Standard routines for calculating the grid points and weights, $w(\theta_i)$, 
for the associated Legendre quadrature formula exist (Press et al., 1992). 
This grid has negligible integration error and also has the slight advantage 
over a purely equiangular grid that the grid points and weights are a 
function of the integration range.  The arrangement of sampled points can 
therefore be made optimal for a sphere with a Galaxy cut. This pixelization 
scheme does have elongated pixels of smaller area near the poles.  Thus, 
since pixel noise scales with pixel area, our scheme contains a 
disproportionately large number of noisy pixels near the poles. In practice, 
we have not found such endpoint apodizing to be harmful. 

What are the optimal choices for $n_\theta $ and $n_\phi $ in terms of 
$l_{max}$? In terms of the Gaussian cubature formula, the optimal choice is 
$n_\theta = l_{max}+1$, $n_\phi = 2l_{max}+1$, which integrates {\it exactly}
the first $(l_{max}+1)^2$ spherical harmonics. The accuracy of this grid is 
easily verified numerically by observing the exact orthonormality of the 
first $(l_{max}+1)^2$ spherical harmonics on the full sky. (Note that 
{\it COBE} type pixelization schemes, which have equal area per pixel, require 
a substantially larger number of points to achieve the same accuracy.)  
Note that because the number of $a_{lm}$'s is half the number of integration 
points, the spherical harmonic transform is {\it not} invertible unless the 
integrated function is bandwidth limited to $l_{max}+1$. Indeed, it has been 
proven (Taylor 1995) that there does not exist {\it any} cubature-based 
discrete spherical harmonic transform with the same number of points as 
spectral coefficients. This highlights an important difference between
Fourier analysis on $R^2$ and on $S^2$. Consider any arbitrary 2D
function $f(\theta,\phi)$. If mapped onto the plane, we can obtain
Fourier coefficients $\hat{f}(l,m)$ with $n_{l}=n_{\theta}$,
$n_{m}=n_{\phi}$, which is invertible. On the sphere, however, the
spherical harmonics by design only span the space of functions
invariant under rotations. This implies symmetries between the $\theta$
and $\phi$ directions, which in the spherical harmonic basis imply
that even though $n_{l}=n_{\theta}$,
$n_{m}=n_{\phi}$, half the terms are missing as $a_{lm}=0$ for $m>l$. This reduction in the number of spectral
coefficients leads to loss of information (by a
factor of 2) for functions which do not respect such symmetries, and
which thus will no longer be invertible. Note that there do not exist
any discrete pixelizations of the sphere which are invariant under the
full group of rotations (for instance, ECP is only azimuthally
symmetric), due to the finite number of Platonic solids.  
   
For practical purposes, the CMB signal is band-width limited, as beam smearing
virtually destroys any signal above some $l_{max}$. However, the noise map 
is not bandwidth limited and cannot be projected onto a finite set of 
spherical harmonics. We see that by noting that the underlying weight map is 
not rotationally invariant (i.e., not statistically isotropic), in
general, which is why the noise matrix ${\bf N}_{(lm)(lm)^{\prime}}$ is dense, even over the full sky. This has
important practical consequences when maximizing the likelihood function. In 
particular, one is not free to transform between pixel and spherical harmonic 
space. It is impossible, for instance, to precondition in spherical harmonic 
space and maximize the likelihood in pixel space: the (essentially noise) 
eigenvalues are wiped out in spherical harmonic space but are retained in 
pixel space. 

We emphasize that our algorithm is independent of the pixelization scheme, 
as long as fast spherical harmonic transforms are available. For instance, 
the HEALPIX scheme (Gorski 1998), which uses equal area pixels, is also a 
viable option.

\section{Spline Fitting}

We have presented an algorithm for extracting the minimum variance power
spectrum, $c_l$, and its error matrix, ${\bf F}^{-1}$, from a map.
This is all that is needed to make comparisons with Gaussian theoretical 
models.  However, there are at least two good reasons why one would like to fit
a smooth curve to the $c_l$ data. 1) We have argued that it is better 
to construct the experimental Fisher matrix from the smoothed $c_{l}$ to avoid
bias in any subsequent analysis. 2) For the purpose of visual presentation 
we would like an aid to guide the eye though a forest of data points.

The usual smoothing procedure is to either bin the data or employ a moving 
average, which reduces the errors by $\sqrt{n}$, where $n$ is the number 
of points in each bin. The drawback of such a procedure is that while the
zeroth- and first-order moments of the data are preserved, higher order
moments are not. In particular, if the underlying function has a non-zero 
second derivative (e.g., at an acoustic peak), a bias is introduced 
(Press et al. 1992).  A significant improvement would be to approximate the 
data as piecewise polynomial, rather than piecewise constant, as with 
Savitzky-Golay smoothing filters (Press et al. 1992).

However, we advocate fitting a least-squares weighted smoothing spline 
(Wahba 1990) to the recovered $c_{l}$. Such a fit is non-parametric and
model-independent in the sense that we allow ourselves $l_{max}-1$ degrees 
of freedom. We find the cubic spline $f(l)$ which minimizes the quantity
\begin{equation}
\sum_{ll^{\prime}} (f(l)-c_l){\bf F}_{ll^{\prime}}(f(l^{\prime})-c_{l^{\prime}})
+ \lambda \int f^{\prime\prime}(l) dl.
\end{equation}
The smoothing parameter $\lambda$, which controls the usual trade-off between 
smoothness and fidelity to the data, is chosen to satisfy the relation
\begin{equation}
(l_{max}-1) - \sqrt{2(l_{max}-1)} \;<\; \chi^2 \;<\; (l_{max}-1) + 
\sqrt{2(l_{max}-1)}
\end{equation}
where $\chi^2 \equiv \sum_{ll^{\prime}} (f(l)-c_l){\bf F}_{ll^{\prime}}
(f(l^{\prime})-c_{l^{\prime}})$. The slight freedom we have in selecting 
$\lambda$ within this range is analogous to the freedom we have to select a 
bin size in an averaging scheme. We proceed with the fit iteratively: first, 
employing the Fisher matrix of the unsmoothed data, we construct a preliminary 
smoothing spline. Next, generating a new Fisher matrix from $f(l)$, we 
construct a second and final smoothing spline. 

An alternative approach to choosing $\lambda$ is to use cross-validation. 
This is essentially a bootstrap technique in which $\lambda$ is chosen so 
that the spline best approximates the data at any given $l$ if it is computed 
with all the data except $c_l$. In general, the two methods give similar 
results. The fits in this paper were obtained using cross-validation.

\begin{figure}
\epsscale{1.00}
\plotone{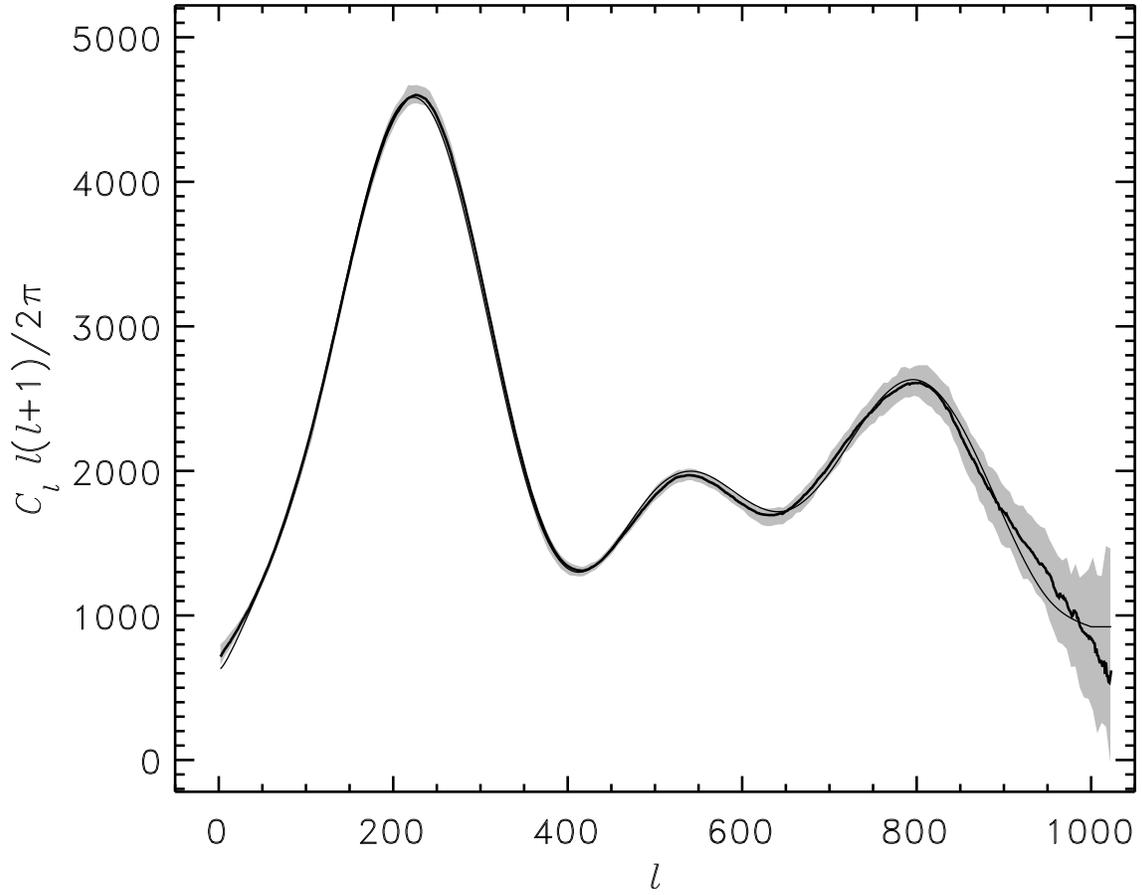}
\caption{The result of a smoothing spline fit to the points in Figure 
\ref{main}. The dark solid line is the spline fit, the light solid line is 
the input power spectrum. The grey band indicates the 68\% confidence band 
obtained from bootstrap simulations. Including the prior expectation that the 
underlying power spectrum is smooth allows one to obtain a very firm handle on 
the power spectrum.}
\label{spline}
\end{figure}

To compute errors on our spline fit, we employ bootstrap resampling. 
Specifically, we calculate the exact Fisher matrix from $f(l)$ and generate 
a series of synthetic power spectra, $c_l^{(i)}$ from this Fisher matrix.
This will be accurate to the extent that we have accurately computed ${\bf F}$. 
One can also generate $c_l^{(i)}$ assuming they are drawn from distributions 
with dispersion $({\bf F}^{-1})_{ll}^{1/2}$, which ignores correlations
between the $c_{l}$. The two methods give very similar results. We then 
compute the smoothing spline for each realization $c_l^{(i)}$ and sort the 
results at each $l$.  This allows us to generate confidence intervals for the 
spline fits.  Figure \ref{spline} shows the customary 68\% confidence band.  

The use of a smoothing spline entails a small but non-negligible bias because 
the mean of the Monte Carlo spline fits does not equal the input power 
spectrum. Such effects are well-known (Wahba 1990). We calibrate this bias 
from the Monte Carlo simulations and apply it to our best fit spline and
confidence bands. Note also that the confidence bands of the spline fit are 
not symmetric about the mean value, particularly near a peak. The use of 
Monte Carlo simulations gives us the probability distribution of the spline 
fits and thus a firm handle on such effects. 

\end{appendix}

\clearpage

\end{document}